\documentclass[apj]{emulateapj}

\usepackage{apjfonts}

\newcommand{\vc}[1]{\textbf{\em #1}}
\newcommand{\bi}[1]{\textbf{\em #1}}

\shorttitle{COMPRESSIBLE MHD TURBULENCE}
\shortauthors{KOWAL \& LAZARIAN}

\begin{document}

\title{Velocity Field of Compressible MHD Turbulence: Wavelet Decomposition and Mode Scalings}

\author{Grzegorz Kowal}
\affil{Department of Astronomy, University of Wisconsin, 475 North Charter Street, Madison, WI 53706, USA}
\affil{Astronomical Observatory, Jagiellonian University, Orla 171, 30-244 Krak\'ow, Poland}
\affil{Instituto de Astronomia, Geof\'\i sica e Ci\^encias Atmosf\'ericas, Universidade de S\~ao Paulo, Rua do Mat\~ao 1226, CEP 05508-900, S\~ao Paulo, Brazil}
\email{kowal@astro.wisc.edu}
\and
\author{Alex Lazarian}
\affil{Department of Astronomy, University of Wisconsin, 475 North Charter Street, Madison, WI 53706, USA}
\email{lazarian@astro.wisc.edu}

\begin{abstract}
We study compressible MHD turbulence, which holds key to many astrophysical
processes, including star formation and cosmic ray propagation.  To account for
the variations of the magnetic field in the strongly turbulent fluid we use
wavelet decomposition of the turbulent velocity field into Alfv\'{e}n, slow and
fast modes, which presents an extension of the Cho \& Lazarian (2003)
decomposition approach based on Fourier transforms.  The wavelets allow to
follow the variations of the local direction of magnetic field and therefore
improve the quality of the decomposition compared to the Fourier transforms
which are done in the mean field reference frame.  For each resulting component
we calculate spectra and two-point statistics such as longitudinal and
transverse structure functions, as well as, higher order intermittency
statistics.  In addition, we perform the Helmholtz-Hodge decomposition of the
velocity field into the incompressible and compressible parts and analyze these
components.  We find that the turbulence intermittency is different for
different components and we show that the intermittency statistics depend on
whether the phenomenon was studied in the global reference frame related to the
mean magnetic field or it was studied in the frame defined by the local magnetic
field.  The dependencies of the measures we obtained are different for different
components of velocity, for instance, we show that while the Alfv\'{e}n mode
intermittency changes marginally with the Mach number the intermittency of the
fast mode is substantially affected by the change.
\end{abstract}

\keywords{ISM: structure --- MHD --- turbulence}

\section{Introduction}
\label{sec:intro}

Astrophysical fluids are magnetized and therefore the astrophysical turbulence
is magnetohydrodynamic (MHD) in its nature.  Compressible MHD turbulence is a
key element for understanding star formation \citep[see][and references
therein]{maclow04a,elmegreen04,mckee07} and velocity fluctuations determine many
of its properties.  For instance, in the modern paradigm of star formation that
turbulent velocity sweep up the matter from large expands of the interstellar
space to create molecular clouds.  Thus, it is important to know the statistical
properties of the velocity field, e.g. its spectrum that reflects how much
energy is associated with the motions at a particular scale (see below).

A further insight into the properties of turbulence, including its generation,
consequences and dissipation calls for the use of more sophisticated measures.
For example, the processes of magnetic field generation depend on the velocity
field vorticity associated with the solenoidal motions, while the processes of
compressing gas are determined by the compressible motions.  In approaching the
problem of decomposing velocity field into solenoidal and dilatational part
following the Helmholtz-Hodge decomposition is frequently attempted
\cite[see][]{federrath08,federrath09}.  Another approach is the decomposition of
the turbulent field in MHD case into basic modes, i.e., Alfv\'en, slow and fast
waves.  While this approach is trivial for the case of the strong magnetic field
with infinitesimal fluctuations \cite[see][]{dobrowolny80}, \cite{cho02a,cho03}
proposed a {\it statistical} decomposition of modes in the Fourier space.  The
statistical nature of the procedure is clear when one considers its application
to strongly perturbed magnetic fields.  As the Fourier transform is defined in
the reference frame related to the mean magnetic field, while the MHD motions
happen in respect to the local magnetic field, there is an inevitable
contribution of all types of motion to the decomposed modes.  However, studying
the cases when the real space decomposition was possible in real space,
\citet[][henceforth CL03]{cho03} showed that the cross-talk between the modes is
small for subAlfv\'enic turbulence.

Testing of the results in CL03 and increasing the accuracy of the MHD mode
decomposition of turbulence is one of the key goals of the present study.  In
doing so in the present paper we make use of the wavelet transformations in
addition to the Fourier transformations.  Wavelets
\cite[see][]{meneveau91a,meneveau91b} present a natural way of describing MHD
turbulence.  Indeed, while in the representation of the \citet[][henceforth
GS95]{goldreich95} model of turbulence the anisotropy is frequently described in
terms of eddies with parallel $k_{\|}$ and perpendicular $k_{\bot}$ wave
vectors, the actual description calls for choosing for $\|$ and $\bot$ in
respect to the {\it local} magnetic field\footnote{Due to this fact that closure
relations used for the model justification in GS95 are doubtful.  The importance
of the {\it local} system of reference was clearly stressed in the works that
followed the original GS95 study
\citep{lazarian99,cho00,maron01,cho02b,cho03b,lithwick01,cho02a}.}.  The latter
is really easy to understand, as it is only local magnetic field that influences
fluid motions at a given point.  Wavelets allow for a local description of the
magnetized turbulent eddies.

In the paper we decompose the turbulent velocity fields using both wavelets and
a more traditional Helmholtz-Hodge decomposition into solenoidal and
compressible parts.  We feel that the latter decomposition is more justified for
the hydrodynamic turbulence than for the MHD turbulence that we study here.
However, we feel that the use of the Helmholtz-Hodge decomposition provides an
additional, although limited, insight into the properties of compressible
motions.

The three major properties of the velocity field that we focus our attention in
the paper are turbulence spectra, anisotropies and intermittency. These three
measures require further description that we provide below.

While turbulence is an extremely complex chaotic non-linear phenomenon, it
allows for a remarkably simple statistical description \cite[see][]{biskamp03}.
If the injections and sinks of the energy are correctly identified, we can
describe turbulence for {\it arbitrary} $Re$ and $Rm$.  The simplest description
of the complex spatial variations of any physical variable, $X({\bf r})$, is
related to the amount of change of $X$ between points separated by a chosen
displacement ${\bf l}$, averaged over the entire volume of interest.  Usually
the result is given in terms of the Fourier transform of this average, with the
displacement ${\bf l}$ being replaced by the wave number ${\bf k}$ parallel to
${\bf l}$ and $|{\bf k}|=1/|{\bf l}|$.  For example, for isotropic turbulence
the kinetic energy spectrum, $E(k)dk$, characterizes how much energy resides at
the interval $k, k+dk$.  At some large scale $L$ (i.e., small $k$), one expects
to observe features reflecting energy injection.  At small scales, energy
dissipation should be seen.  Between these two scales we expect to see a
self-similar power-law scaling reflecting the process of non-linear energy
transfer.  We shall attempt to get the power-law scalings for the components of
the velocity field.

The presence of a magnetic field makes MHD turbulence anisotropic
\citep{montgomery81,matthaeus83,shebalin83,higdon84,goldreich95}\cite[see][for
review]{oughton03}.  The relative importance of hydrodynamic and magnetic forces
changes with scale, so the anisotropy of MHD turbulence does too.  Many
astrophysical results, e.g. the dynamics of dust, scattering and acceleration of
energetic particles, thermal conduction, can be obtained if the turbulence
spectrum and its anisotropy are known \cite[see][for review]{lazarian09b}.  The
knowledge of the anisotropy of Alfvenic mode of MHD turbulence determines the
extend of magnetic field wandering influencing heat transfer
\citep{narayan01,lazarian06a} and magnetic reconnection
\citep{lazarian99,kowal09}.

We would like to stress that in what follows we discuss the properties of {\it
strong} MHD turbulence.  This type of turbulence is not directly related to the
amplitude of the magnetic perturbations, however.  The low-amplitude turbulence
can be strong and isotropically driven turbulence with $\delta B\ll B$ at the
injection scale exhibits only a limited range of scales for which it is weak
\cite[see][]{galtier00}, while at sufficiently small scales it gets strong
\cite[see the discussion in][]{lazarian99}.

An anisotropic spectrum alone, say $E({\bf k)}\,d{\bf k}$, cannot characterize
MHD turbulence in all its complexity because it involves only the averaged
energy in motions along a particular direction.  To have a full statistical
description, one needs to know not only the averaged spectrum of a physical
variable but higher orders as well.  The tendency of fluctuations to become
relatively more violent but increasingly sparse in time and space as the scales
decreases, so that their influence remains appreciable, is called {\it
intermittency}.  The intermittency increases with the ratio of the size scales
of injection and dissipation of energy, so the very limited range of scales
within numerical simulations may fail to reflect the actual small scale
processes.  The turbulence intermittency can result in an important intermittent
heating of the interstellar medium \citep{falgarone05,falgarone06,falgarone07}.

In this article we investigate the scaling properties of the structure functions
of velocity and its components for compressible MHD turbulence with different
sonic and Alfv\'{e}nic Mach numbers.  In \S\ref{sec:models} we describe the
numerical models of compressible MHD turbulence.  We decompose velocity into a
set of components including the incompressible and compressible parts, and MHD
waves: Alfv\'{e}n, slow and fast using methods described in \S\ref{sec:decomp}.
In \S\ref{sec:spectra} we study spectra of velocity and its components.  In
\S\ref{sec:anisotropy} we study the anisotropy of dissipative structures.  We
show differences in the structures for different components.  In
\S\ref{sec:intermittency} we study the scaling exponents and the intermittency
of velocity structures.  We show their dependence on the sonic and Alfv\'{e}n
regime of turbulence.  In \S\ref{sec:discussion} we discuss our results and
their relation to the previous studies.  In \S\ref{sec:summary} we draw our
conclusions.

\section{Numerical simulations}
\label{sec:models}

We used an second-order-accurate essentially nonoscillatory (ENO) scheme
\citep[see][]{cho02a} to solve the ideal isothermal MHD equations in a periodic
box,
\begin{eqnarray}
 \frac{\partial \rho}{\partial t} + \nabla \cdot (\rho \bi{v}) = 0, \\
 \frac{\partial \rho \bi{v}}{\partial t} + \nabla \cdot \left[ \rho \bi{v} \bi{v} + \left( a^2 \rho + \frac{B^2}{8 \pi} \right) \bi{I} - \frac{1}{4 \pi}\bi{B}\bi{B} \right] = \bi{f},  \\
 \frac{\partial \bi{B}}{\partial t} - \nabla \times (\bi{v} \times \bi{B}) = 0,
 \label{eq:induction}
\end{eqnarray}
where $\rho$ is density, $\bi{v}$ is velocity, $\bi{B}$ is magnetic field, and
$a$ is the isothermal speed of sound.  We incorporated the field interpolated
constrained transport (CT) scheme \citep[see, e.g.,][]{toth00} into the
integration of Eq.~(\ref{eq:induction}) to maintain the $\nabla \cdot \bi{B} =
0$ constraint numerically.  On the right-hand side, the source term $\bi{f}$
represents a random solenoidal large-scale driving force.  The rms velocity
$\delta v$ is maintained to be approximately unity, so that $\bi{v}$ can be
viewed as the velocity measured in units of the rms velocity of the system and
$\bi{B}/\left(4 \pi \rho\right)^{1/2}$ as the Alfv\'{e}n velocity in the same
units.  The time $t$ is in units of the large eddy turnover time ($\sim L/\delta
v$) and the length in units of $L$, the scale of the energy injection.  The
magnetic field consists of the uniform background field and a fluctuating field:
$\bi{B}= \bi{B}_\mathrm{ext} + \bi{b}$.  Initially $\bi{b}=0$.  We use units in
which the Alfv\'{e}n speed $v_A=B_\mathrm{ext}/\left(4 \pi \rho\right)^{1/2}=1$
and $\rho=1$ initially.  The values of $B_\mathrm{ext}$ have been chosen to be
similar to those observed in the ISM turbulence.

\begin{deluxetable*}{cccccccccc} 
\tablewidth{0pt}
\tablecolumns{10}
\tabletypesize{\footnotesize}
\tablecaption{List of all analyzed models.\label{tab:models}}
\tablehead{
 \colhead{model} &
 \colhead{$B_\mathrm{ext}$} &
 \colhead{$P_\mathrm{ini}$} &
 \colhead{${\cal M}_{s}$} &
 \colhead{${\cal M}_{A}$} &
 \colhead{res.} &
 \colhead{max time} &
 \colhead{max $\delta \rho$} &
 \colhead{max $\delta V$} }
\startdata
B1P1    & 1.0 & 1.0  & 0.68$^{\pm 0.03}$ & 0.68$^{\pm 0.03}$ & $512^3$ & 7.0 & $\sim$0.3  & $\sim$0.6 \\
B1P.1   & 1.0 & 0.1  & 2.20$^{\pm 0.09}$ & 0.69$^{\pm 0.03}$ & $512^3$ & 7.0 & $\sim$0.9  & $\sim$0.7 \\
B1P.01  & 1.0 & 0.01 & 7.0$^{\pm 0.3}$   & 0.70$^{\pm 0.04}$ & $512^3$ & 7.0 & $\sim$1.9  & $\sim$0.7 \\
\\
B.1P1   & 0.1 & 1.0  & 0.74$^{\pm 0.06}$ & 7.4$^{\pm 0.6}$ & $512^3$ & 7.0 & $\sim$0.2  & $\sim$0.7 \\
B.1P.1  & 0.1 & 0.1  & 2.34$^{\pm 0.08}$ & 7.4$^{\pm 0.3}$ & $512^3$ & 7.0 & $\sim$0.8  & $\sim$0.7 \\
B.1P.01 & 0.1 & 0.01 & 7.1$^{\pm 0.3}$   & 7.1$^{\pm 0.3}$ & $512^3$ & 7.0 & $\sim$1.5  & $\sim$0.7 \\
\enddata
\end{deluxetable*}

For our calculations, similar to our earlier studies \citep{kowal07a}, we
assumed that $B_\mathrm{ext}/\left(4 \pi \rho\right)^{1/2} \sim \delta B/\left(4
\pi \rho\right)^{1/2} \sim \delta v$. In this case, the sound speed is the
controlling parameter, and basically two regimes can exist: supersonic and
subsonic.  Note that within our model, supersonic means low $\beta$, i.e. the
magnetic pressure dominates, and subsonic means high $\beta$, i.e. the gas
pressure dominates.

We present 3D numerical experiments of compressible MHD turbulence for a broad
range of Mach numbers ($0.2\le{\cal M}_s\le7.1$ and ${\cal M}_A\sim 0.2$ or
$\sim1.8$; see Table \ref{tab:models}).  The model name contains two letters:
``P'' and ``B'' followed by a number.  The letters B and P mean the external
magnetic field and the initial gas pressure, respectively, and the numbers
designate the value of the corresponding quantity.  For example, a name
``B.1P.01'' points to an experiment with $B_\mathrm{ext}=0.1$ and $P=0.01$. We
understand the Mach number to be defined as the mean value of the ratio of the
absolute value of the local velocity $v$ to the local value of the
characteristic speed $c_s$ or $v_A$ (for the sonic and Alfv\'{e}nic Mach number,
respectively).

We drove the turbulence at wave scale $k$ equal to about 2.5 (2.5 times smaller
than the size of the box).  This scale defines the injection scale in our
models.  We did not set the viscosity and diffusion explicitly in our models.
The scale at which the dissipation starts to act is defined by the numerical
diffusivity of the scheme.  The ENO-type schemes are considered to be relatively
low diffusion ones \cite[see, e.g.][]{liu98,levy99}.  The numerical diffusion
depends not only on the adopted numerical scheme but also on the ``smoothness''
of the solution, so it changes locally in the system.  In addition, it is a
time-varying quantity.  All these problems make its estimation very difficult
and incomparable between different applications.  However, the dissipation
scales can be estimated approximately from the velocity spectra.  In the case of
our models we estimated the dissipation scale $k_{\nu}$ at 22 for the resolution
256$^3$.

\section{Decomposition of the velocity field into components}
\label{sec:decomp}

\subsection{Compressible and incompressible parts}

Using the Hodge generalization of the Helmholtz theorem we can split an
arbitrary vector field $\bi{u}$ into three components:
\begin{equation}
 \bi{u} = \bi{u}_p + \bi{u}_s + \bi{u}_l ,
\end{equation}
where each component has specific properties:
\begin{itemize}
 \item[a)] Potential component ($\bi{u}_p$) - it is curl-free component, i.e.
$\nabla \times \bi{u}_p = 0$, so it stems from a scalar potential $\phi$:
\begin{equation}
 \bi{u}_p = \nabla \phi.
\end{equation}
 The scalar potential $\phi$ is not unique. It is defined up to a constant. This
component describes the compressible part of the velocity field.

 \item[b)] Solenoidal component ($\bi{u}_s$) - it is divergence-free component,
i.e. $\nabla \cdot \bi{u}_s = 0$, so it stems from a vector potential ${\bf
\Phi}$:
\begin{equation}
 \bi{u}_s = \nabla \times {\bf \Phi}.
\end{equation}
 The vector potential ${\bf \Phi}$ also is not unique. It is defined only up to
a gradient field. In the case of velocity this component describes the
incompressible part of the field.

 \item[c)] Laplace component (${\bf u}_l$) - it is both divergence-free and
curl-free. Laplace component comes from a scalar potential which satisfies the
Laplace differential equation $\Delta \phi = 0$.
\end{itemize}

Thus the decomposition can be rewritten in the form:
\begin{equation}
 \bi{u} = \nabla \times {\bf \Phi} + \nabla \phi + \bi{u}_l .
 \label{eqn:decompose}
\end{equation}

Applying the divergence operation on both sides and using the divergence-free
property of the solenoidal and Laplace components we obtain
\begin{equation}
 \nabla \cdot \bi{u} = \nabla^2 \phi = \Delta \phi.
\end{equation}
To find the scalar potential $\phi$ and the potential field $\bi{u}_p$ we have
to solve Poisson equation for $\nabla \cdot \bi{u}$.

To calculate the vector potential ${\bf \Phi}$ we apply curl operation on both
sides of the Eq.~(\ref{eqn:decompose}). Similarly, using the divergence-free
property of the potential and Laplace fields we results in equation:
\begin{equation}
 \nabla \times \bi{u} = \nabla \times \nabla \times {\bf \Phi} = \Delta {\bf \Phi}.
\end{equation}
Here, the calculation of the vector potential ${\bf \Phi}$ requires to solve
triple set of Poisson equations, one equation for each component of ${\bf
\Phi}$.

The simulations with periodic boundary conditions have the advantage, that we
can solve the Poisson equation using Fourier methods. The Fourier components of
the velocity field are transformed back to the real space then, and further
analyzed.

\begin{figure*} 
 \epsscale{0.8}
 \plotone{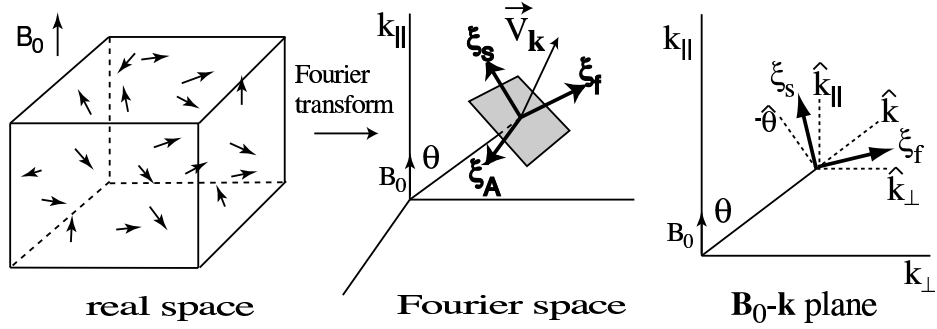}
 \caption{Graphical representation of the mode separation method. We separate
the Afv\'{e}n, slow and fast modes by the projection of the velocity Fourier
component ${\bf v}_k$ on the bases ${\bf \hat\xi}_A$, ${\bf \hat\xi}_s$ and
${\bf \hat\xi}_f$, respectively. Figure taken from CL03 \label{fig:separation}}
\end{figure*}

\subsection{Separation into the Alfv\'{e}n, Slow and Fast modes}

Another very important type of decomposition is the separation of velocity into
the MHD waves: Alfv\'{e}n, slow and fast.  In this paper, we use an extended
mode based on a technique described in CL03.  The procedure of decomposition is
performed in the Fourier space by a simple projection of the velocity Fourier
components $\hat{\bi{u}}$ on the direction of the displacement vector for each
mode (see \ref{fig:separation}).  The directions of the displacement vectors
${\bf \hat\xi}_s$, ${\bf \hat\xi}_f$, and ${\bf \hat\xi}_A$ corresponding to the
slow mode, fast and Alfv\'{e}n modes, respectively, are defined by their unit
vectors
\begin{equation}
 {\bf \hat\xi}_s \propto (-1 + \alpha - \sqrt{D}) k_\parallel {\bf \hat{k}_\parallel} + (1 + \alpha - \sqrt{D}) k_\perp {\bf \hat{k}_\perp} \, ,
\end{equation}
\begin{equation}
 {\bf \hat\xi}_f \propto (-1 + \alpha + \sqrt{D}) k_\parallel {\bf \hat{k}_\parallel} + (1 + \alpha + \sqrt{D}) k_\perp {\bf \hat{k}_\perp} \, ,
\end{equation}
\begin{equation}
 {\bf \hat\xi}_A = - {\bf \hat\varphi} = {\bf \hat{k}}_\perp \times {\bf \hat{k}}_\parallel \, ,
\end{equation}
where ${\bf k}_\parallel$ and ${\bf k}_\perp$ are the parallel and perpendicular
to ${\bf B}_\mathrm{ext}$ components of wave vector, respectively, $D = (1 +
\alpha)^2 - 4 \alpha \cos^2 \theta$, $\alpha = a^2 / V_A^2$, $\theta$ is the
angle between ${\bf k}$ and ${\bf B}_\mathrm{ext}$, and ${\bf \hat\varphi}$ is
the azimuthal basis in the spherical polar coordinate system.  The Fourier
components of each mode can be directly used to calculate spectra.  For other
measures, such as structure functions, we transform them back to the real space.

We extend the CL03 technique by introducing an additional step before the
Fourier separation, in which we decompose each component of the velocity field
into orthogonal wavelets using discrete wavelet transform
\citep[see][e.g.]{antoine99}
\begin{equation}
 \vc{U}(a, \vc{w}_{lmn}) = a^{-N/2} \sum_{\vc{x}_{ijk}}{\psi\left( \frac{\vc{x}_{ijk} - \vc{w}_{lmn}}{a} \right) \vc{u}(\vc{x}_{ijk}) \Delta^N \vc{x}},
\end{equation}
where $\vc{x}_{ijk}$ and $\vc{w}_{lnm}$ are $N$-dimensional position and
translation vectors, respectively, $a$ is the scaling parameter,
$\vc{u}(\vc{x}_{ijk})$ is the velocity vector field in the real space,
$\vc{U}(\vc{x}_{ijk})$ is the velocity vector field in the wavelet space, and
$\psi$ is the ortogonal analysing function called wavelet.  The sum in the
equation is taken over all position indices.  We use 12-tap Daubechies wavelet
as an analyzing function and fast discrete version of the wavelet transform
\citep{antoine99}, thus, as a result we obtain a finite number of wavelet
coefficients.  After the wavelet transform of the velocity we calculate the
Fourier representation of each wavelet coefficient and perform its individual
separation into the MHD waves in the Fourier space using the CL03 method and
then update the Fourier coefficients of all MHD waves iterating over all
wavelets.  In this way we obtain a Fourier representation of the Alfv\'en, slow
and fast waves.  The final step is the inverse Fourier transform all all wave
components.

This additional step allows for important extension of the CL03 method, namely,
allows for the local definition of the mean magnetic field and density used to
calculate $\alpha$ and $D$ coefficients.  Since the individual wavelets are
defined locally both in the real and Fourier spaces, the averaging of the mean
field and density is done only within the space of each wavelet.

We can summarize the extended version of the MHD waves decomposition into the
following steps:
\begin{enumerate}
 \item Perform wavelet transform of all velocity components.
 \item Iterate over all wavelet coefficients.
 \begin{enumerate}
  \item Calculate Fourier representation of the current wavelet.
  \item Calculate the mean density and magnetic field within the space occupied
by the wavelet.
  \item Perform the CL03 separation in the Fourier space.
  \item Add the contribution from the separated wavelet to the Fourier
representation of the Alfv\'en, slow and fast modes.
 \end{enumerate}
 \item Perform the inverse Fourier transform of the Alfv\'en, slow and fast wave
components.
\end{enumerate}

\section{Spectra of the Velocity Components}
\label{sec:spectra}

In order to obtain one-dimensional (1D) spectra we first calculate Fourier
transform of a quantity and next multiply it by its conjunction.  We can use this
procedure because we used a fully periodic domain. The 3D spectra must be
averaged or integrated over shells $k_n \le k < k_{n+1}$. We used a simple
integration by summing all squared amplitudes at given shell. For each model we
have collected data several time steps. We used them to increase the size of
sample and to measure the time departures of the spectra from their mean
profiles. We should note here, however, that the time averaging and standard
deviations were calculated in log-log space. Otherwise, we could result in
taking logarithm of negative number, e.g. for ranges of $k$ where the power
spectrum has very small values (of order 10$^{-6}$ or less).

\begin{figure*}  
 \epsscale{0.48}
 \plotone{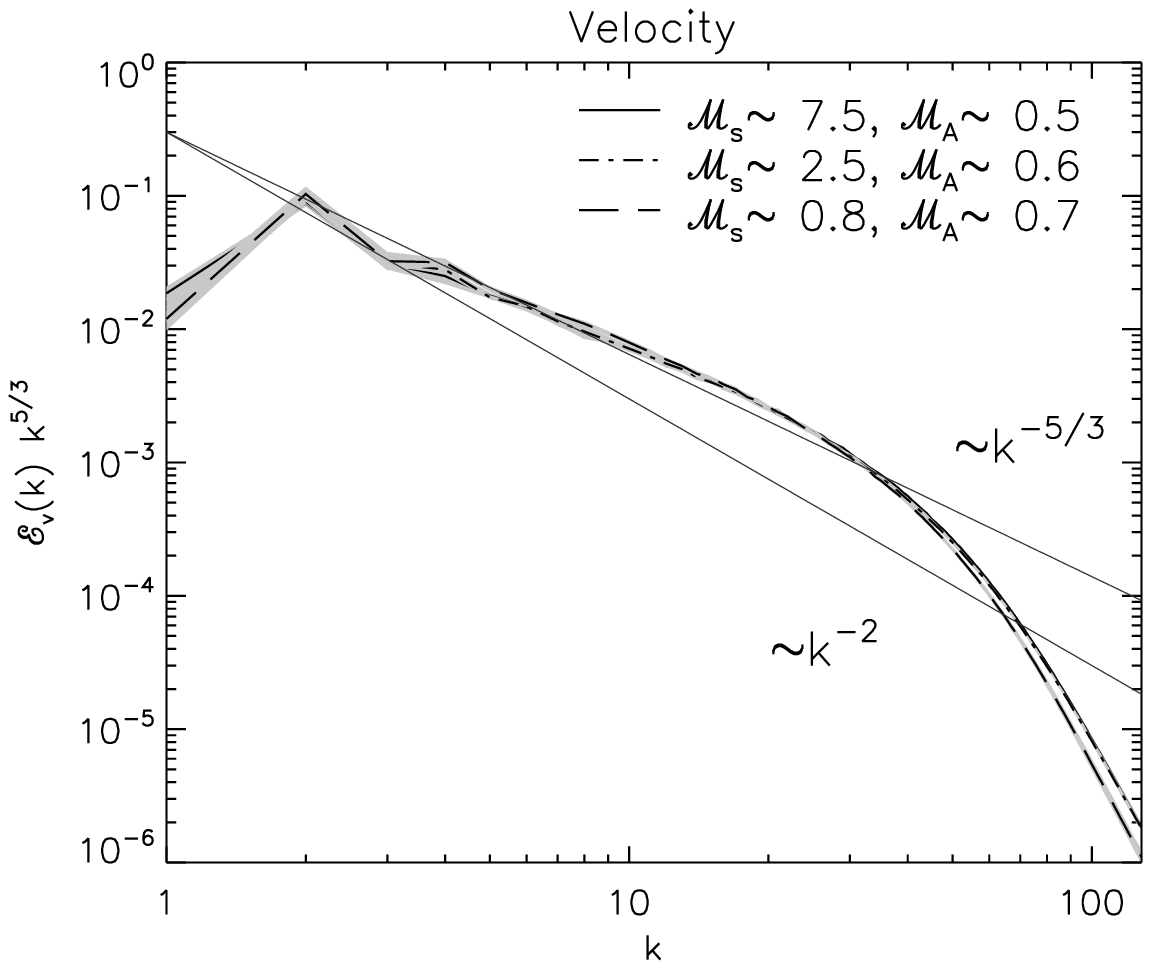}
 \plotone{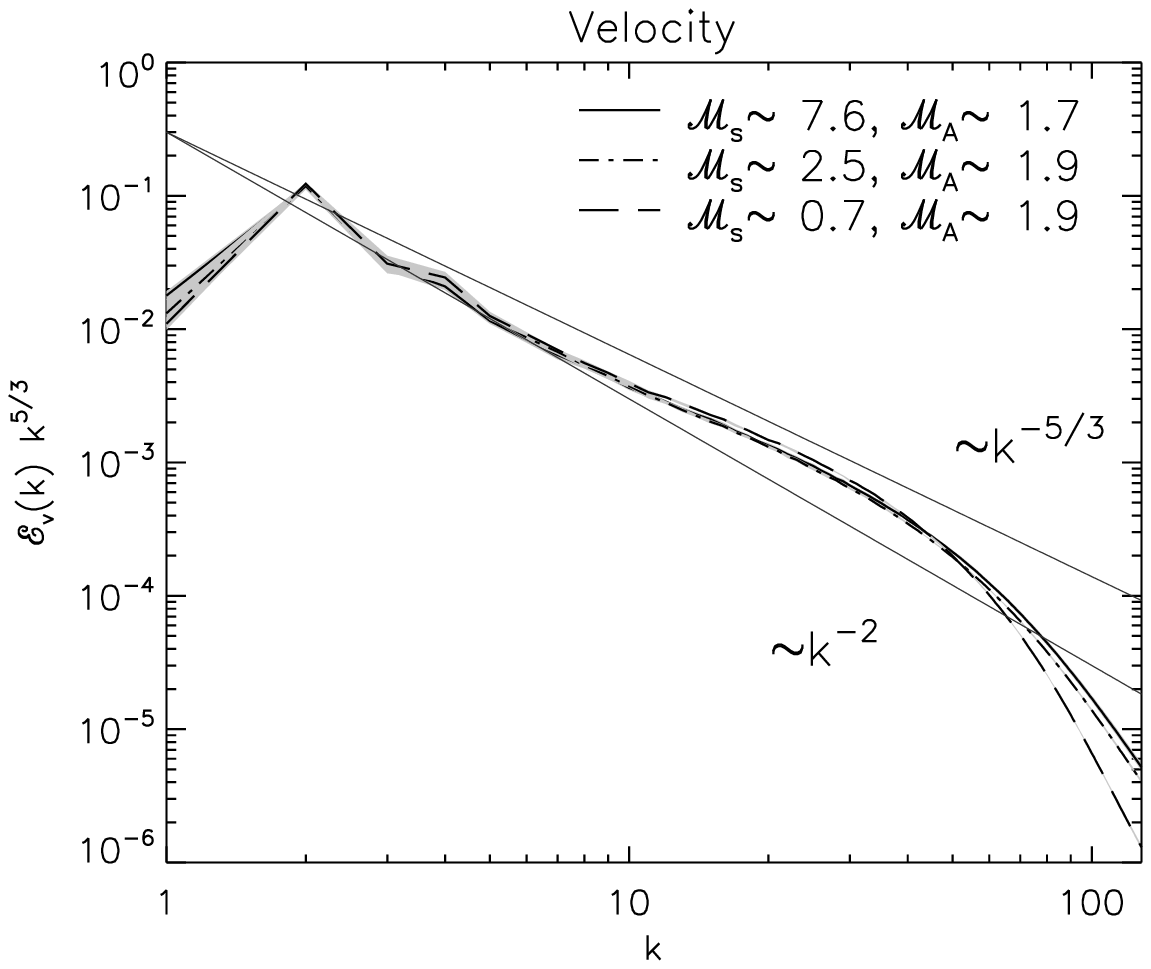}
 \plotone{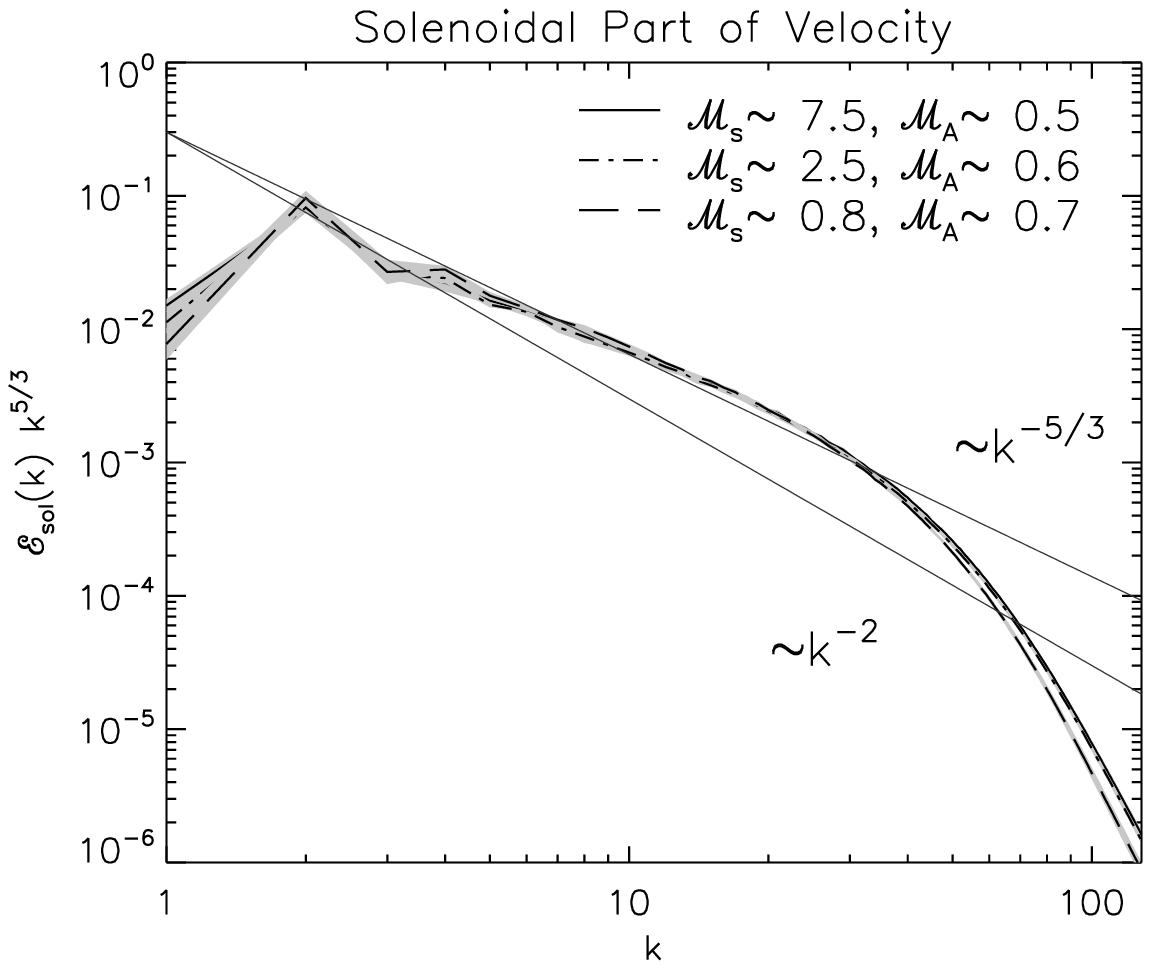}
 \plotone{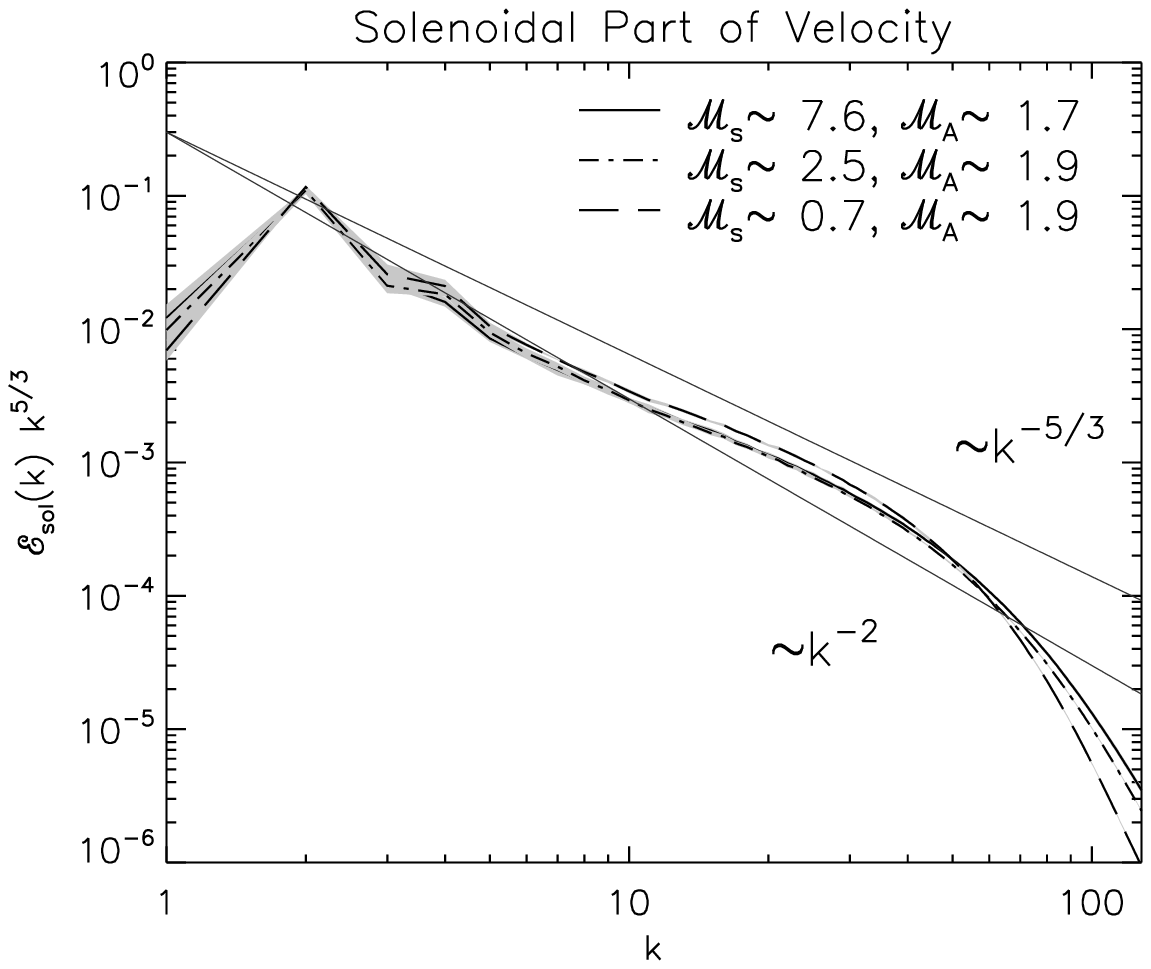}
 \plotone{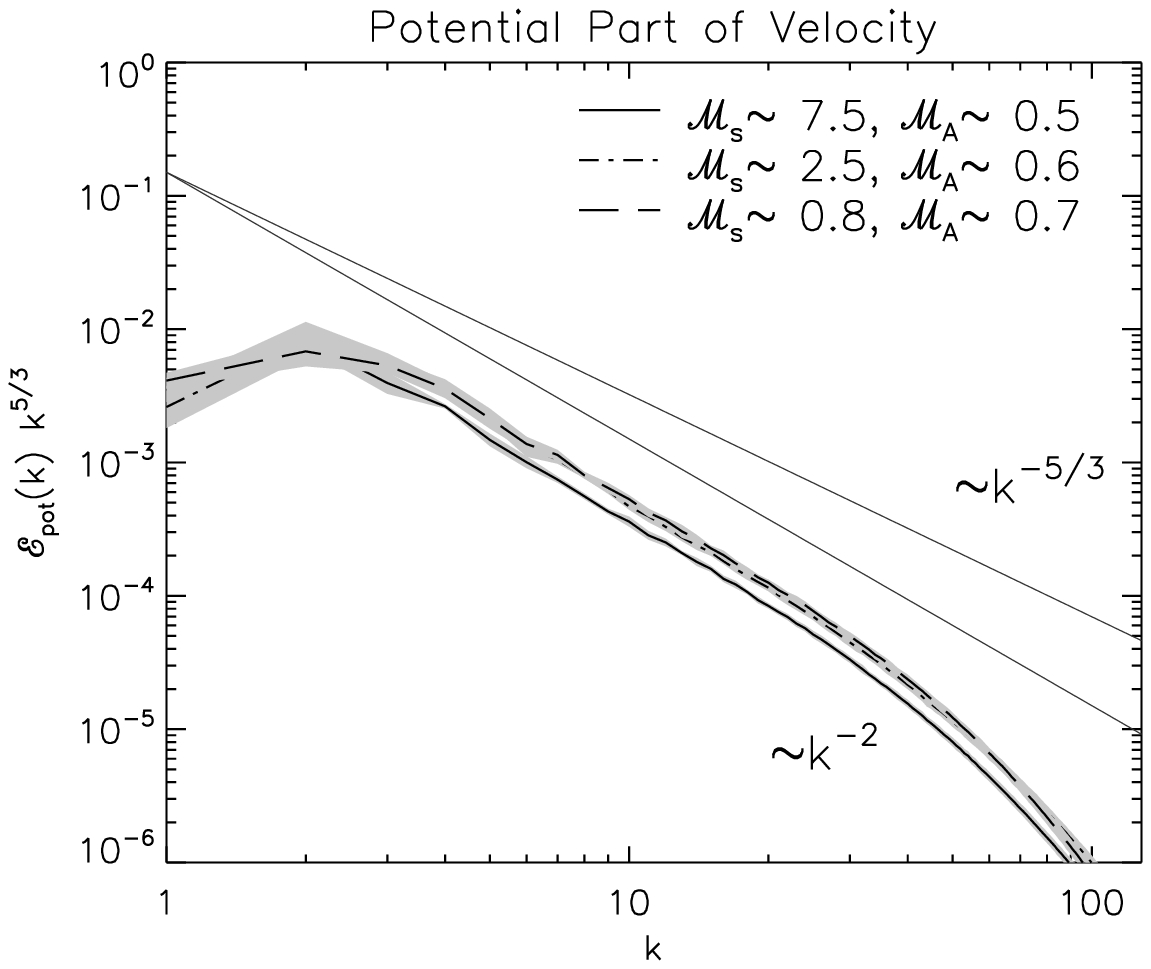}
 \plotone{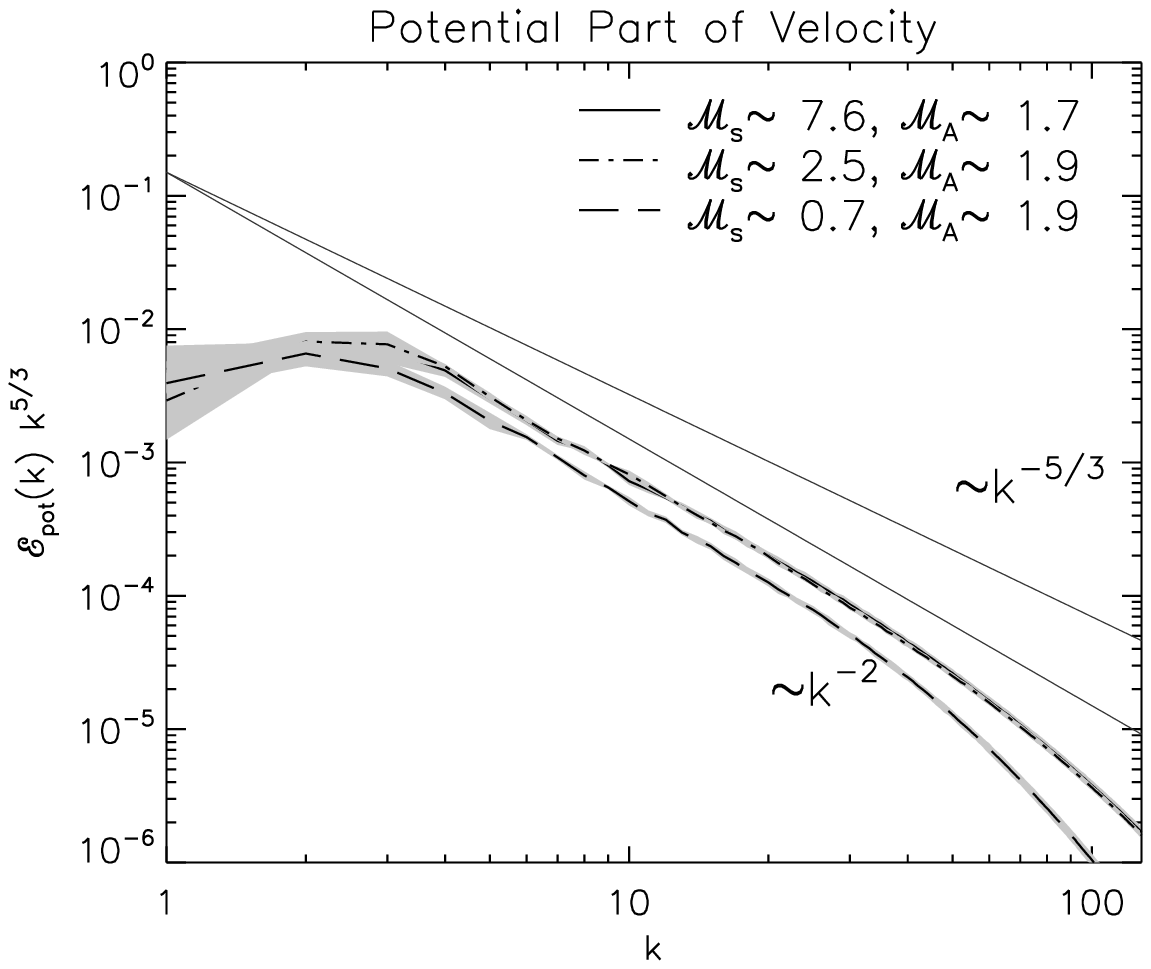}
 \caption{Spectra of the velocity field (top row) and its solenoidal and potential parts (middle and bottom rows, respectively) for experiments with different sonic Mach numbers for two magnetic regimes: subAlfv\'{e}nic (left column) and superAlfv\'{e}nic (right column). The gray area denotes the degree of time variance of the spectra. \label{fig:velo_spectra}}
\end{figure*}

In Figure~\ref{fig:velo_spectra} we present spectra for velocity field (top row)
and its incompressible and compressible parts (middle and bottom rows,
respectively) for models with different sonic Mach numbers. The left column in
Figure \ref{fig:velo_spectra} shows models with a strong magnetic field (${\cal
M}_A \sim 0.5-0.7$), while the right column shows the corresponding models but
with a weak magnetic field (${\cal M}_A \sim 1.7-1.9$). The spectra are averaged
over several time snapshots taken for a fully developed turbulence.  This
allowed us to estimate the spectra time variance which is shown as the gray area
plotted around the mean profiles.  We see that in all cases those variances are
small.

Within the inertial range, which we estimated to be $k \in [ 4, 20 ]$, the
spectra of velocity field slightly change their spectral indices with the value
of ${\cal M}_s$.  In the case of subAlfv\'{e}nic models the spectral index is
close to $-5/3$.  In the case of superAlfv\'{e}nic turbulence, the indices do
not change with ${\cal M}_s$ as well, and all models show the same, close to
$-2$, value of the spectral index at low wave numbers, which is reflection of
the presence of shocks in the system.  Proceeding toward smaller scales we
observe a bump with spectral index approaching $-5/3$ which could be explained
by the growing importance of the mean field at these scales and dominance of
Alfv\'en wave.  The velocity fluctuations have comparable amplitudes for
corresponding scales for all models.  It means, that the strength of
fluctuations at particular scale $k$ within the inertial range does not depend
on  ${\cal M}_s$.

The total velocity field contains two components: solenoidal, which is
equivalent to the incompressible part, and potential, which contains the
compressible part of the field and the remaining part which is curl- and
divergence-free.  In Table~\ref{tab:energies} we show the percentage
contribution of each component to the total velocity field.  We see, that the
compressible part constitutes only a fraction of the total field.  However, the
magnitude of this fraction is different for sub- and supersonic models.  In the
case of subAlfv\'{e}nic turbulence it is about 3\% in subsonic models and about
7\% in supersonic models, which confirms a higher efficiency of the compression
in the presence of supersonic flows.  Furthermore, the fraction also changes
when we compare models with strong and weak magnetic fields.  The velocity
field, in the presence of a weak magnetic field, contains about 5\% of the
compressible part in the model with ${\cal M}_s \sim 0.7$ and even up to 16\% in
models with ${\cal M}_s>1$.  The consequence of the presence of a strong
magnetic field results in a reduction of the compressible part of the velocity
field by a factor 2.  This indicates a substantial role of the magnetic field in
the damping of the generation of the compressible flows.

\begin{deluxetable}{cc|cc|ccc} 
\tablewidth{0pt}
\tablecolumns{9}
\tablecaption{Percentage amount of the kinetic energy contained within each velocity component.  Errors correspond to a measure of the time variation.\label{tab:energies}}
\tablehead{
 \colhead{${\cal M}_{s}$} &
 \colhead{${\cal M}_{A}$} &
 \colhead{$V_\mathrm{incomp.}$} &
 \colhead{$V_\mathrm{comp.}$} &
 \colhead{$V_A$} &
 \colhead{$V_s$} &
 \colhead{$V_f$} }
\startdata
  $\sim 0.7$ & $\sim 0.7$ & 96.5$^{\pm0.8}$ &  3.3$^{\pm0.8}$ & 58$^{\pm4}$ & 37$^{\pm3}$ & 4.8$^{\pm0.7}$ \\
  $\sim 2.2$ & $\sim 0.7$ & 93$^{\pm2}$     &  7$^{\pm2}$     & 58$^{\pm5}$ & 33$^{\pm4}$ & 9$^{\pm2}$ \\
  $\sim 7.0$ & $\sim 0.7$ & 92$^{\pm2}$     &  7$^{\pm2}$     & 56$^{\pm4}$ & 36$^{\pm4}$ & 8.0$^{\pm0.7}$ \\
\\
  $\sim 0.7$ & $\sim 7.4$ & 95$^{\pm2}$ & 5$^{\pm2}$  & 52$^{\pm4}$ & 42$^{\pm4}$ & 6.2$^{\pm0.8}$ \\
  $\sim 2.3$ & $\sim 7.4$ & 86$^{\pm1}$ & 14$^{\pm2}$ & 47$^{\pm3}$ & 37$^{\pm4}$ & 16$^{\pm2}$    \\
  $\sim 7.1$ & $\sim 7.1$ & 84$^{\pm2}$ & 16$^{\pm2}$ & 47$^{\pm4}$ & 33$^{\pm4}$ & 20$^{\pm2}$
\enddata
\end{deluxetable}

Due to a substantial dominance of the incompressible part of the velocity field,
we expect that its spectra should follow the spectra for the total velocity
field.  In the middle row of Figure~\ref{fig:velo_spectra} we present the
spectra for the incompressible part of the velocity field.  We see that the
spectra, at least within the inertial range, are very similar to those observed
for the velocity field.  Also the spectral indices have very close values for
models with strong and weak magnetic field.

In the case of compressible turbulence, more interesting are spectra of the
potential component (the bottom row in Figure~\ref{fig:velo_spectra}).  We see
that these spectra are different from those for the incompressible part.  The
strength of the compressible part confirms the change of contribution of the
compressible part.  It is smaller for subsonic models than for supersonic
models, which can be seen in the bottom row of Figure~\ref{fig:velo_spectra}.
However, both supersonic models have almost the same spectra with the spectral
indices about $-2$ for subAlfv\'{e}nic turbulence and for superalfv\'{e}nic
turbulence.  Both supersonic spectra show almost exactly the same profiles and
amplitudes at all scales, even within the dissipation range.  This means that
the amount of compressible part of the velocity field do not change with ${\cal
M}_s$ when its value is larger than one.

The second decomposition, more important for MHD turbulence, separates the
velocity field into three different MHD waves: an incompressible Alfv\'{e}n and
slow and fast magneto acoustic waves, which both are compressible.  In
Table~\ref{tab:energies} we included the percentage amount of these components
in the total velocity field.  As we see, the most of energy is contained in the
Alfv\'{e}n wave.  It is almost 60\% in the case of subAlfv\'{e}n turbulence, and
about 50\% for superAlfv\'{e}nic turbulence.  The slow wave contains
approximately 1/3 of the total energy.  However, for superAlfv\'{e}nic case,
this amount is slightly higher.  Table~\ref{tab:energies} suggests, that the
slow wave is weaker when the turbulence become supersonic.  We do not see
similar behavior for the Alfv\'{e}n wave in the case of models with a strong
magnetic field.  This effect could take place also in the superAlfv\'{e}nic
models, but it is weakened by relatively large errors.  An interesting
dependence is observed in the case of the fast wave.  Although, the fast wave is
the weakest among all MHD waves, it strongly depends on the regime of
turbulence.  Similarly to the compressible part of velocity field, it is
stronger for models with weak magnetic field.  In addition, it is much stronger
when turbulence are supersonic, but this strength seems to be weakly dependent
on the sonic Mach number.

\begin{figure*}  
 \epsscale{0.48}
 \plotone{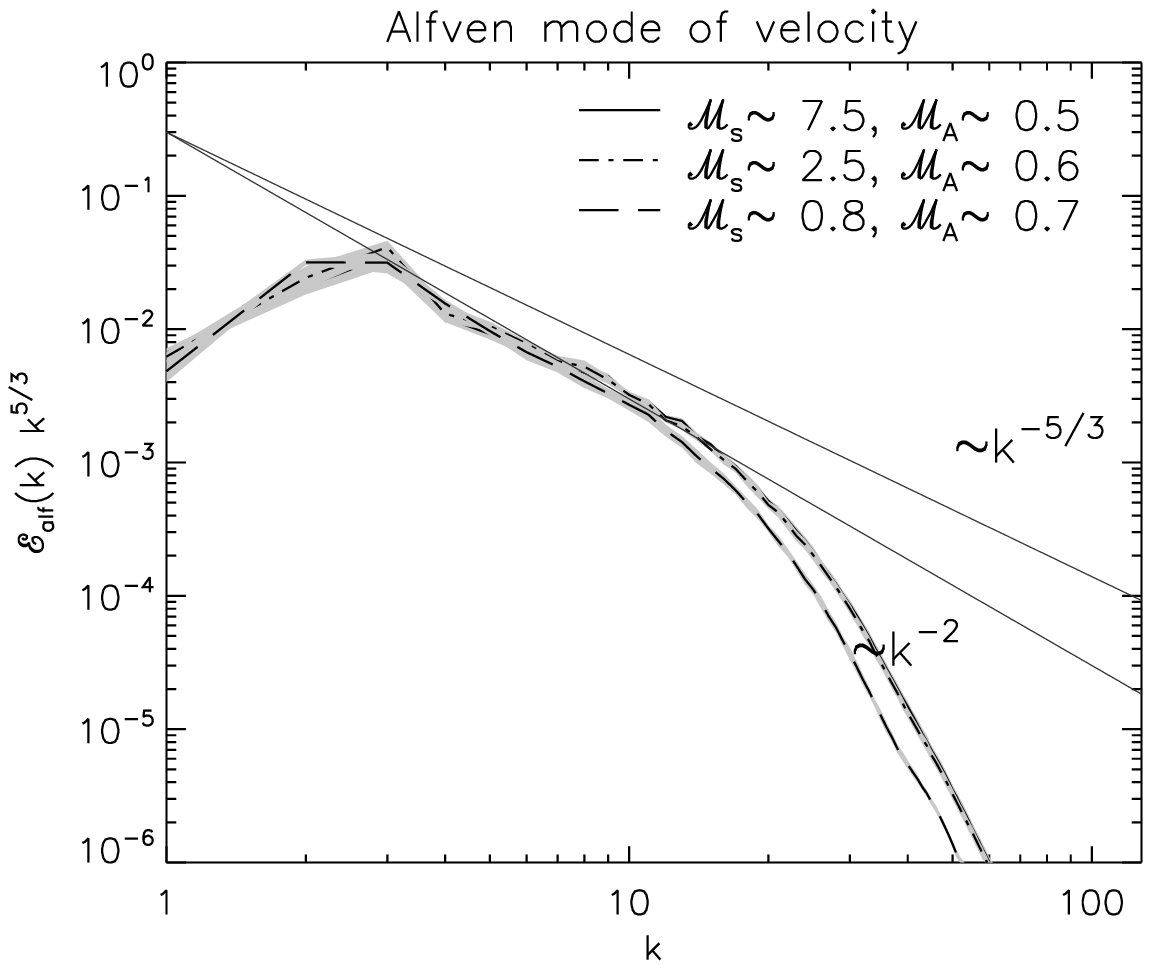}
 \plotone{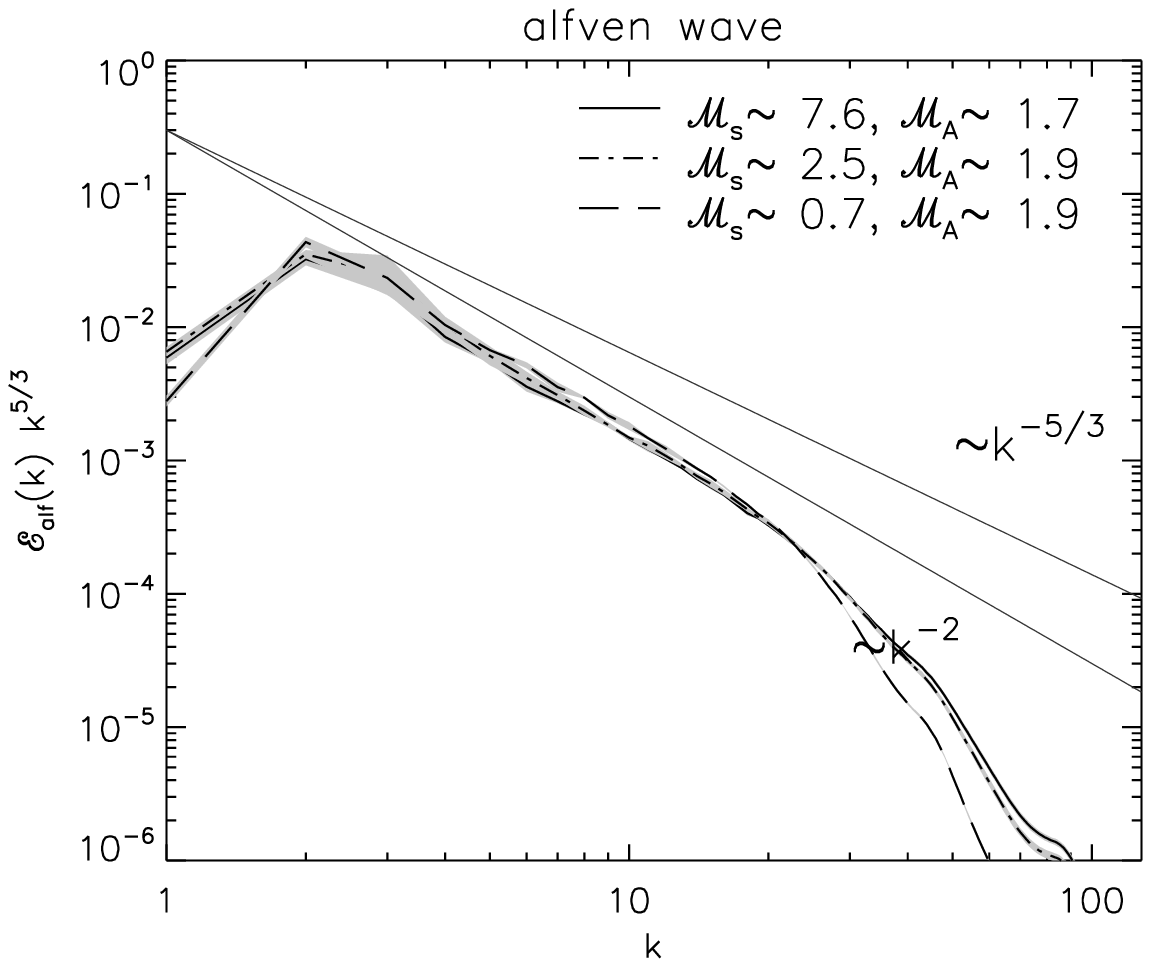}
 \plotone{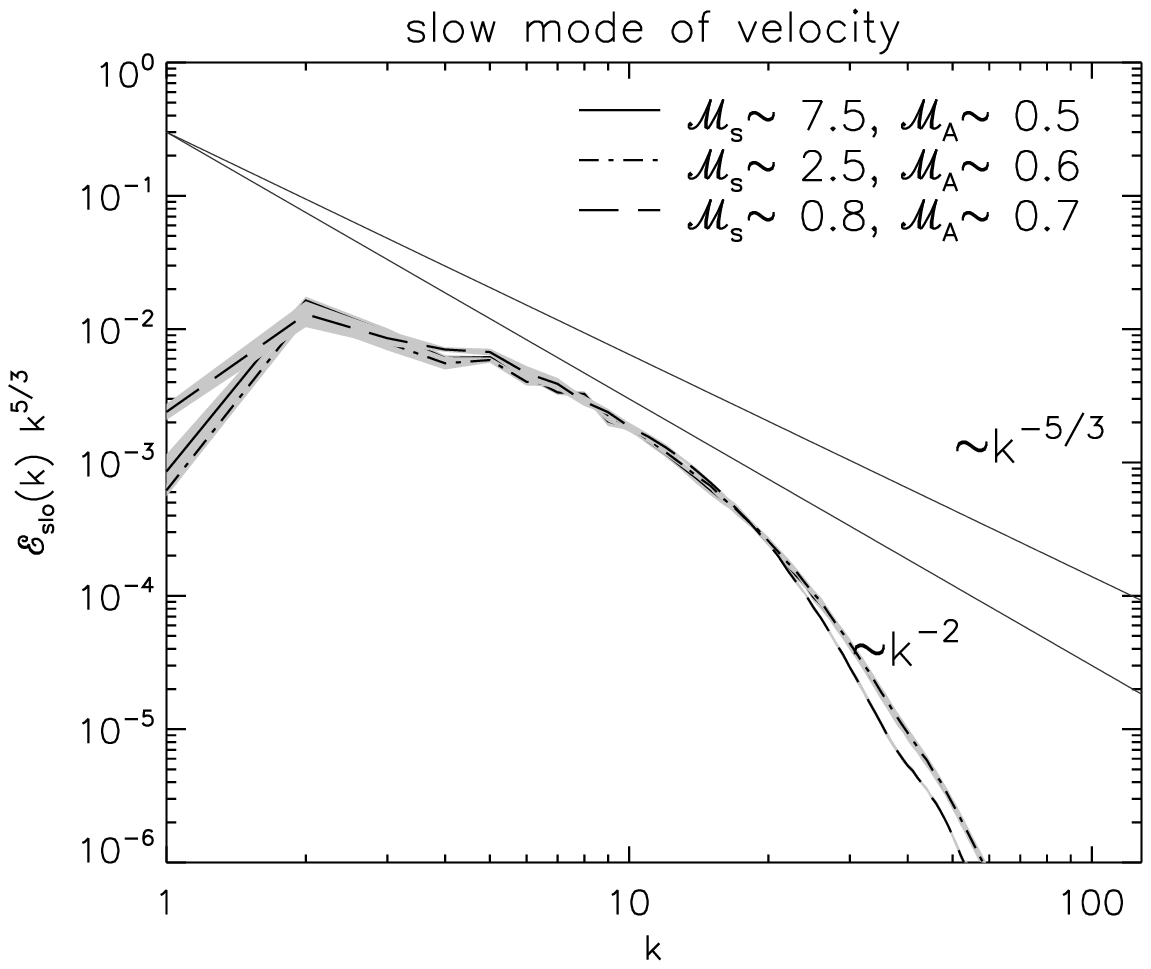}
 \plotone{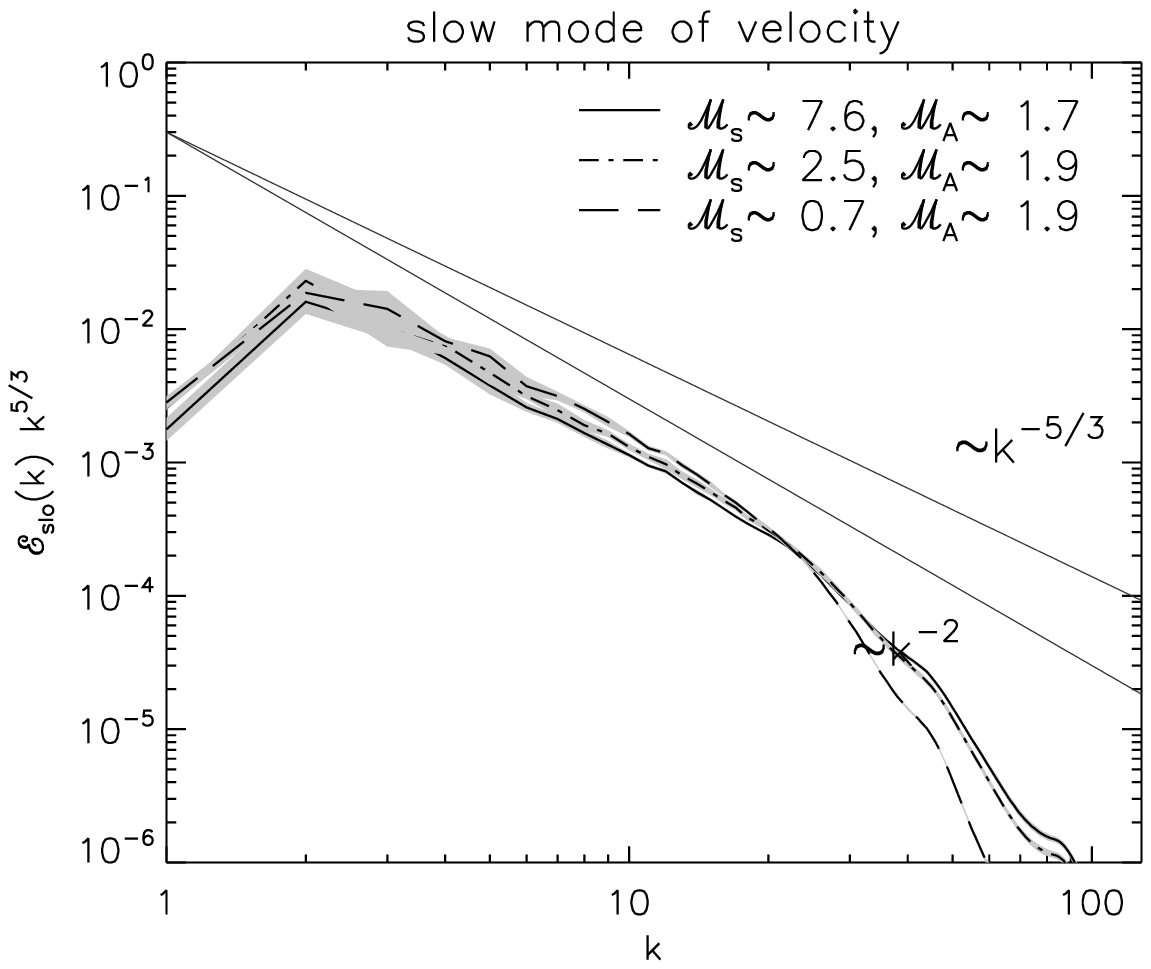}
 \plotone{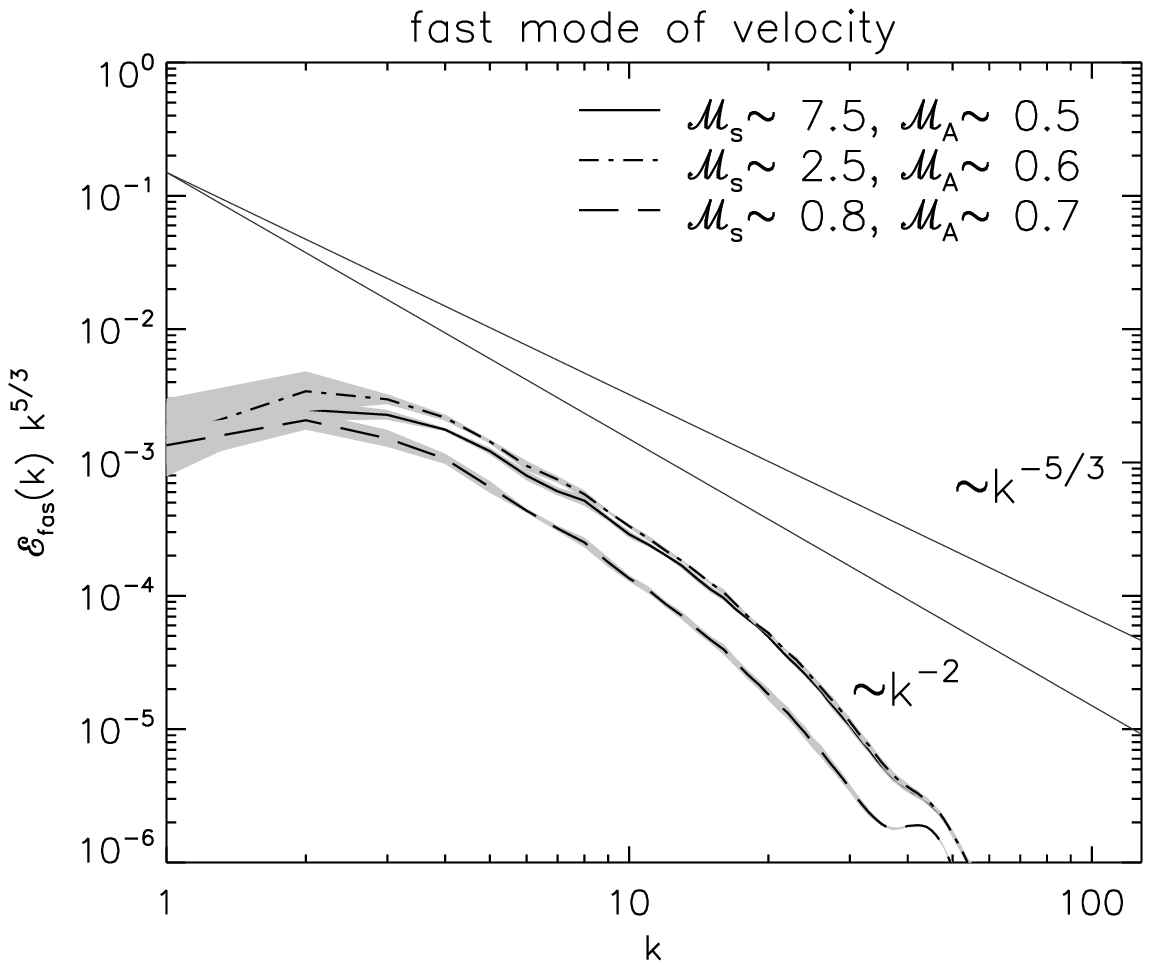}
 \plotone{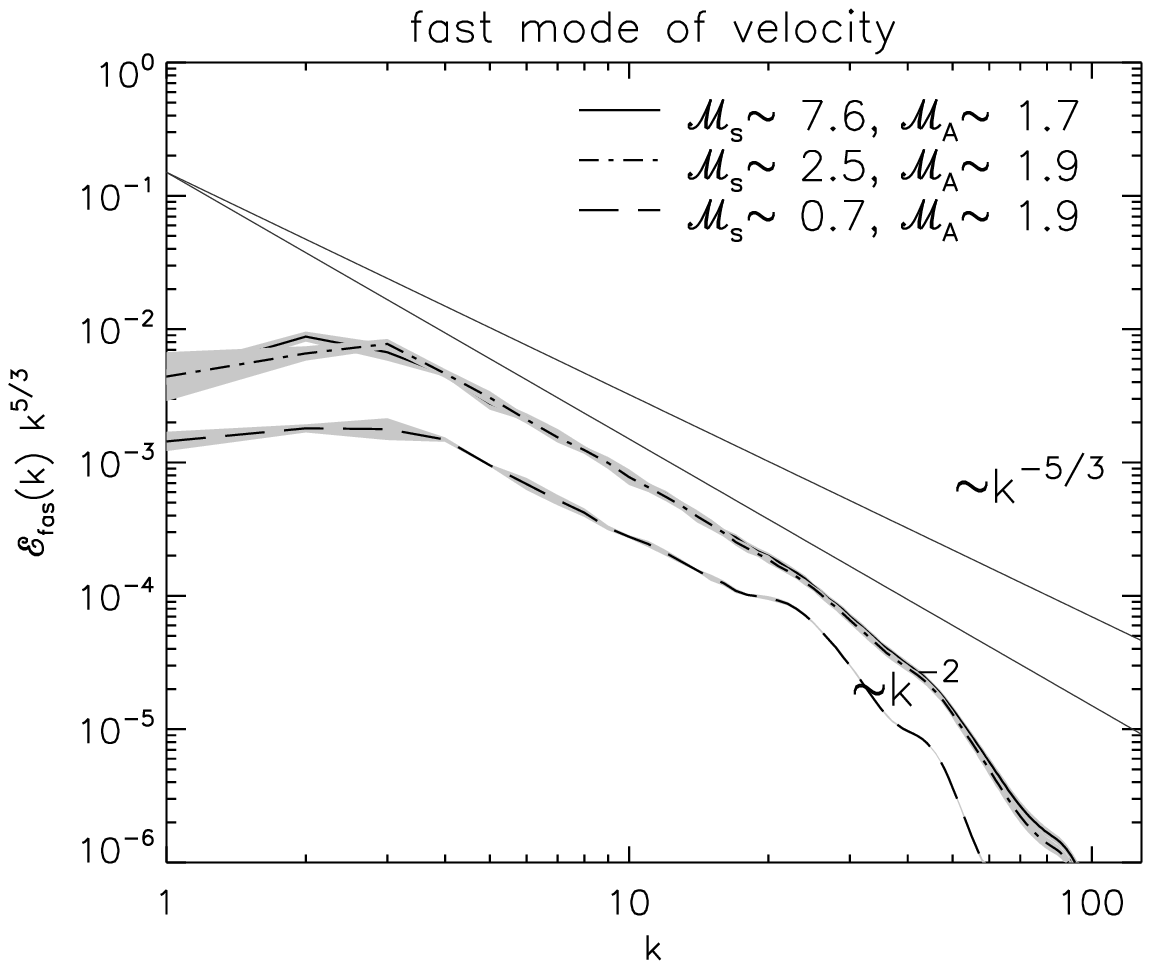}
 \caption{Spectra of the Alfv\'{e}n, slow and fast mode (top, middle and bottom rows, respectively) of the velocity field for experiments with different sonic Mach numbers in two regimes: subAlfv\'{e}nic (left column) and superAlfv\'{e}nic (right column). The gray area denotes the degree of departures at single time snapshots from the mean profile. \label{fig:spectra_waves}}
\end{figure*}

In Figure~\ref{fig:spectra_waves} we present spectra for the MHD waves for all
models from Table~\ref{tab:models}.  In the left column we show spectra for
models with a strong magnetic field, while in the right column spectra for
models with a weak magnetic field.  Top, middle and bottom rows show the
Alfv\'{e}n, slow and fast mode spectra, respectively.  In the case of Alfv\'{e}n
wave we see that the spectral indices depend on the sonic Mach number weakly.
All indices lay between $-2$ and $-5/3$ with a slight dependence on ${\cal
M}_s$.  This is similar situation like for spectra of velocity field.  For
superAlfv\'{e}nic models we observe similar situation, however, since the mean
field is weaker, the power spectra are shifted down to smaller amplitudes. The
similarities in spectra between the Alfv\'{e}n mode and velocity field are due
to the fact, that the Alfv\'{e}n mode constitutes the major part of the velocity
field.

The slow wave, however, constitutes also a substantial part of the velocity
field.  Looking closer in the spectral indices for the slow mode plotted in
Figure~\ref{fig:spectra_waves} we see that their values do not change
significantly with ${\cal M}_s$ when mean magnetic field is strong, although it
is difficult to determine one spectral index do to changing profile of the
spectrum with the wave number.  There seems to be stronger dependence of the
spectral indices on the sonic Mach number when the mean field is weak, where the
inertial region is easier to determine.

The fast wave, the weakest mode of the velocity field, shows two types of
spectra depending on the sonic and Alfv\'{e}nic regime of turbulence.  In the
case of subsonic models we see that the fast mode spectrum has index close to
$-2$.  When the field is weak, the value of index is a bit flatter then $-2$.
This indicates a clear dissimilarity of spectra for subsonic turbulence, for
different strength of the magnetic field.  In supersonic models, however, the
spectral indices are going closer to the value $-2$, independently of the sonic
Mach number.  This signifies a growing role of the shock compression in
supersonic turbulence.

\section{Anisotropy}
\label{sec:anisotropy}

If we want to study the anisotropy of turbulent structures we need to introduce
a reference frame.  In the case of magnetized turbulence the reference frame is
defined in the natural way by the local mean magnetic field.  The local magnetic
field is computed using the procedure of smoothing by a three-dimensional
Gaussian profile with the width equal to the separation length.  Since the
volume of smoothing grows with the separation length $l$, the direction of local
mean magnetic field might change with $l$ at arbitrary point.  This is an
extension of procedures employed in \cite{cho00} and \cite{cho02b}.

To analyze the anisotropy of different components of the velocity field we use
the total structure function
\begin{equation}
 S^{(p)}(l)=\langle \delta v ({\bf l}) ^p \rangle,
 \label{eqn:str_func}
\end{equation}
where $\delta v$ is a total increment
\begin{equation}
 \delta v({\bf l}) = | {\bf v}({\bf x} + {\bf l}) - {\bf v}({\bf x}) |,
 \label{eqn:vel_incr}
\end{equation}
and $\langle ... \rangle$ denotes an ensemble averaging.  For each component we
evaluate the parallel and perpendicular structure functions taking respectively
the directions of the separation length ${\bf l}$ parallel and perpendicular to
the local mean magnetic field.  To show the degree of anisotropy we plot the
structure function in direction perpendicular to the local mean magnetic field
as a function of the parallel structure function for corresponding separation
lengths.  In this way we show also, how the anisotropy changes with the scale of
structures.

\begin{figure*}  
 \epsscale{0.45}
 \plotone{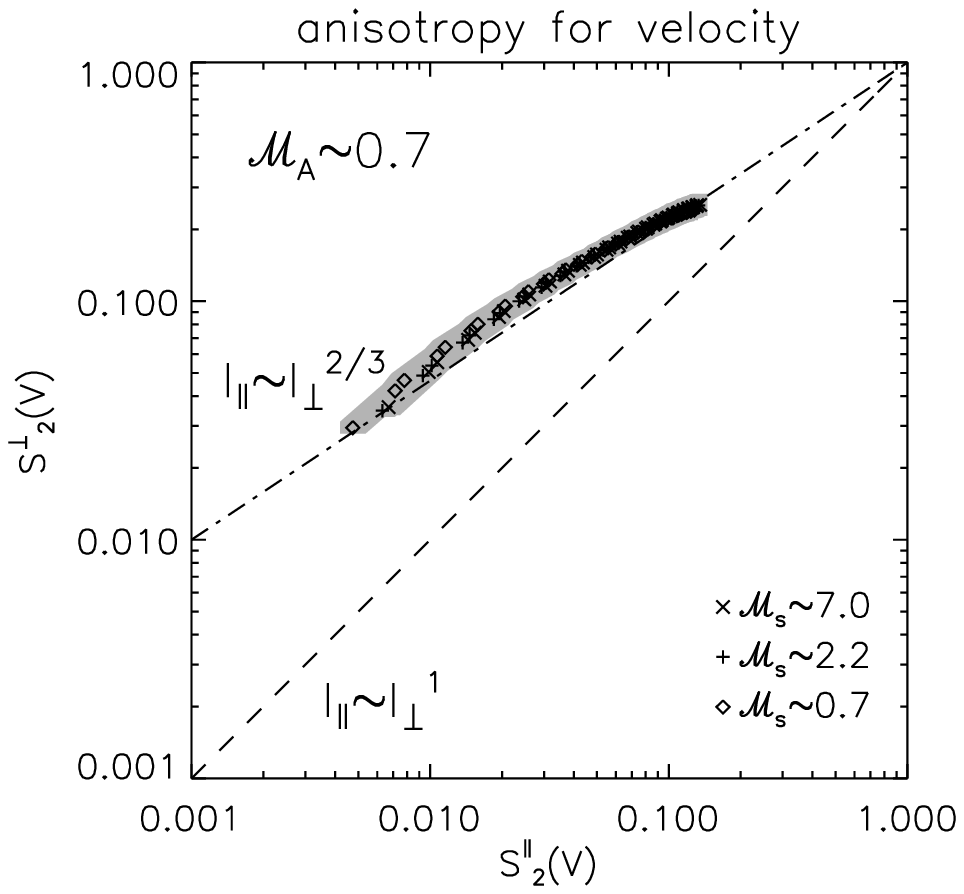}
 \plotone{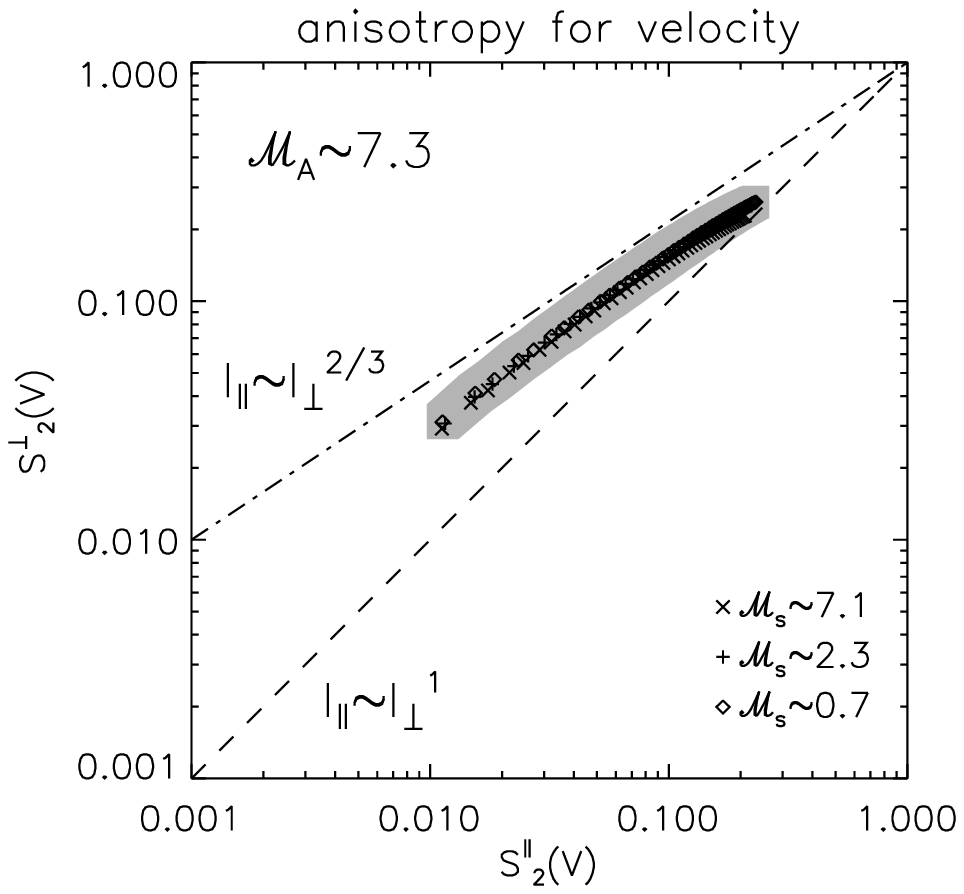}
 \plotone{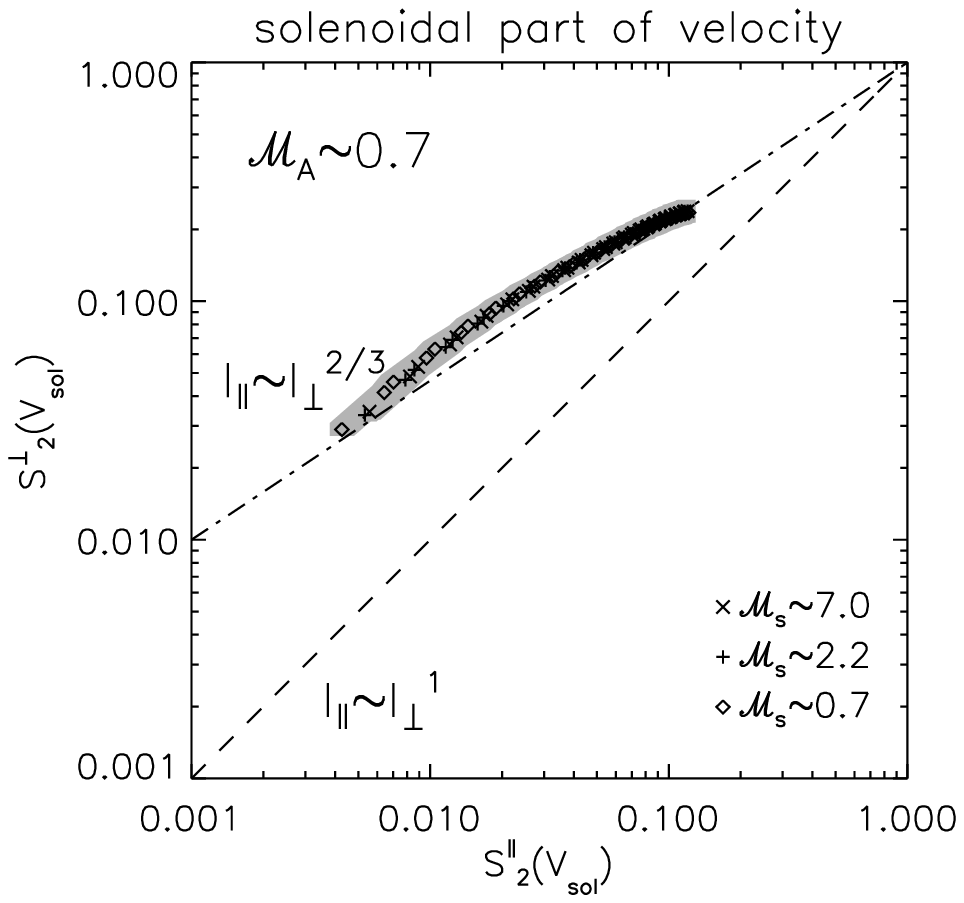}
 \plotone{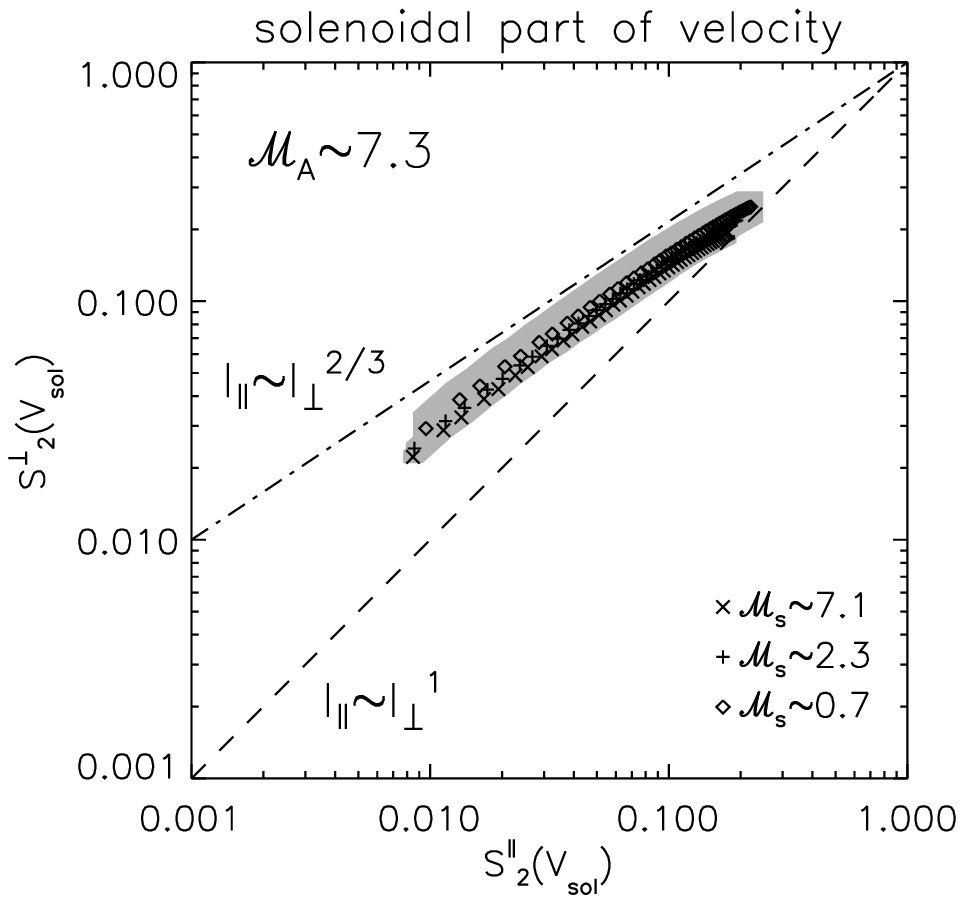}
 \plotone{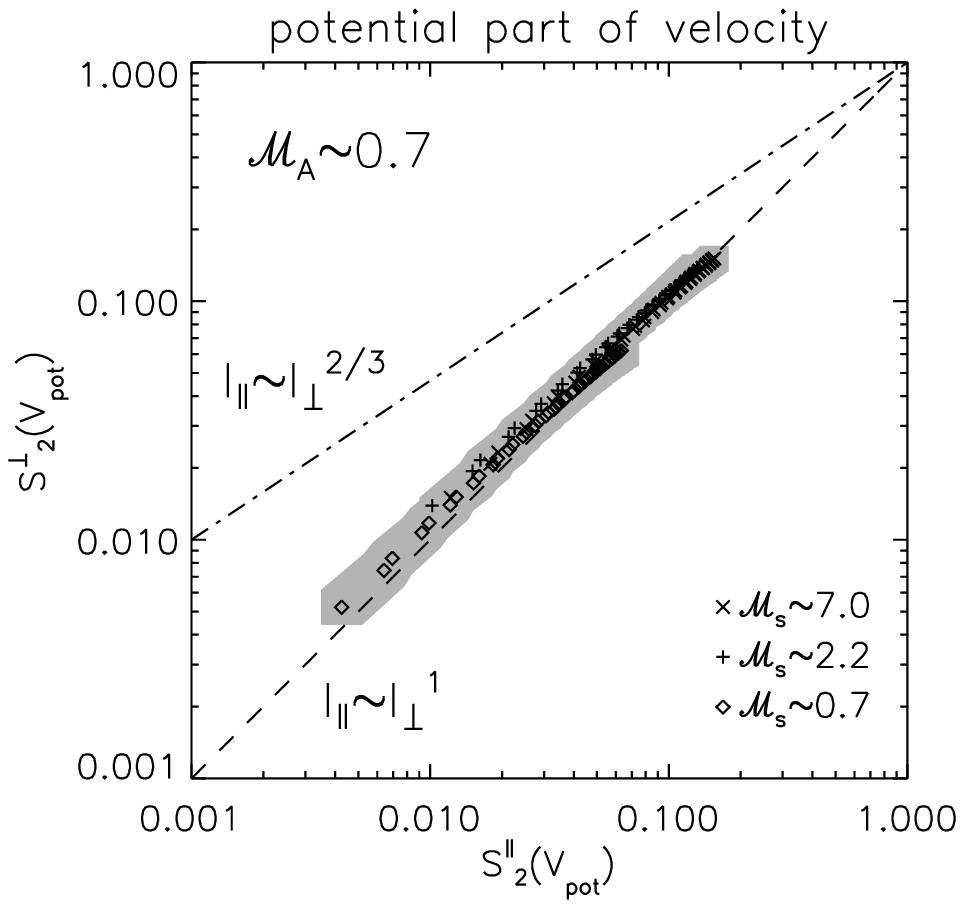}
 \plotone{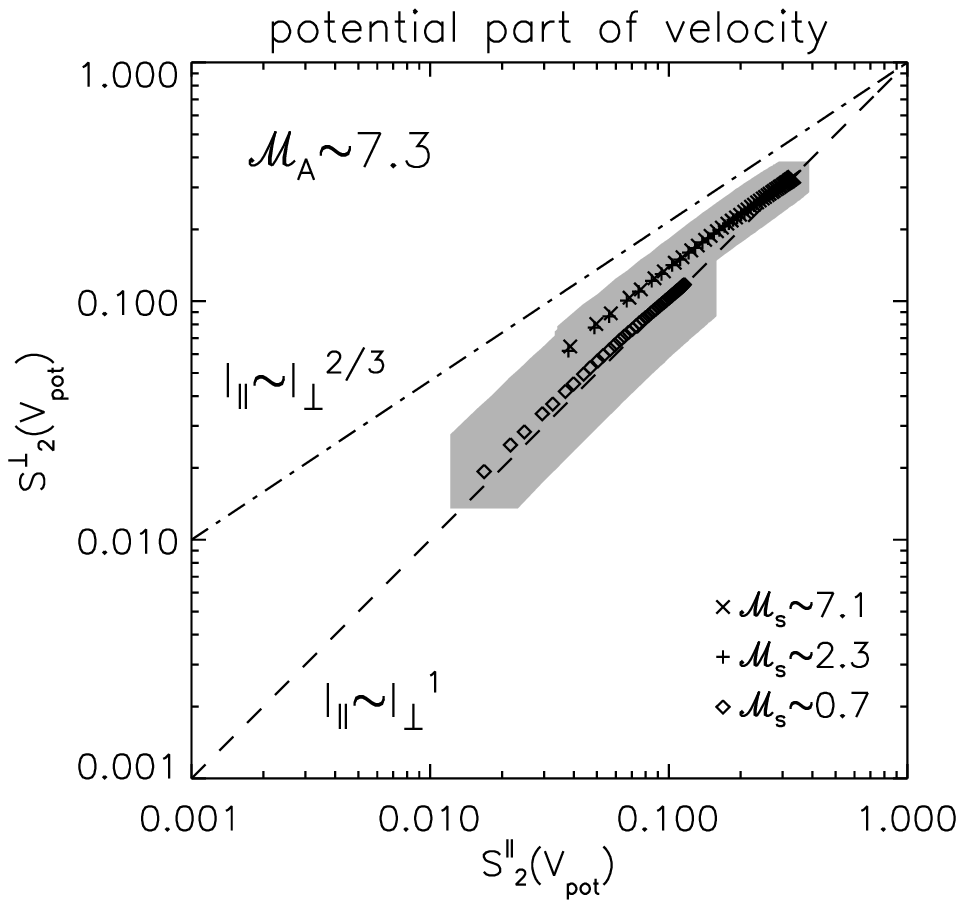}
 \caption{Anisotropy of the structures of the velocity (top row) and solenoidal and potential parts of the velocity (middle and bottom rows, respectively). The left column shows anisotropies for subAlfv\'{e}nic models, the right column plots are for superAlfv\'{e}nic models. To show anisotropy we use the 2$^\mathrm{nd}$-order total structure functions, parallel and perpendicular to the local mean magnetic field. Points corresponds to the mean profiles of the structure functions averaged over several snapshots. The gray areas under points correspond to the degree of departures of the structure functions in time. \label{fig:comp_anisotropy}}
\end{figure*}

\begin{figure*}  
 \epsscale{0.45}
 \plotone{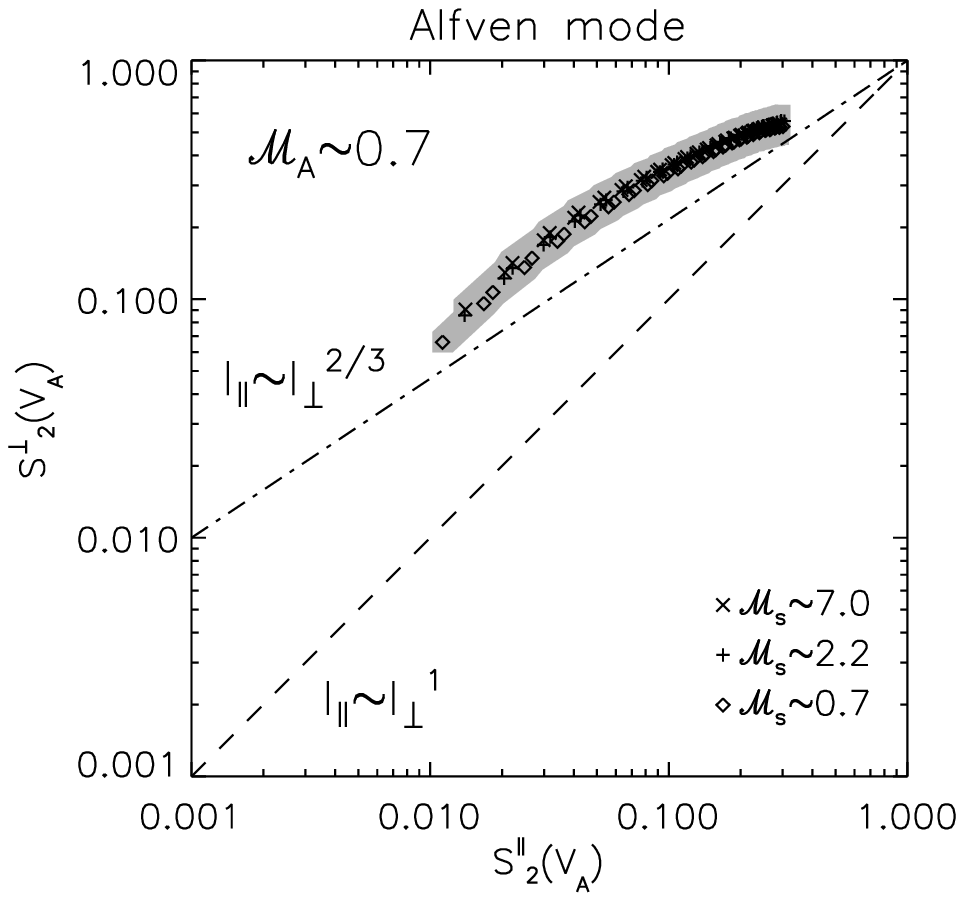}
 \plotone{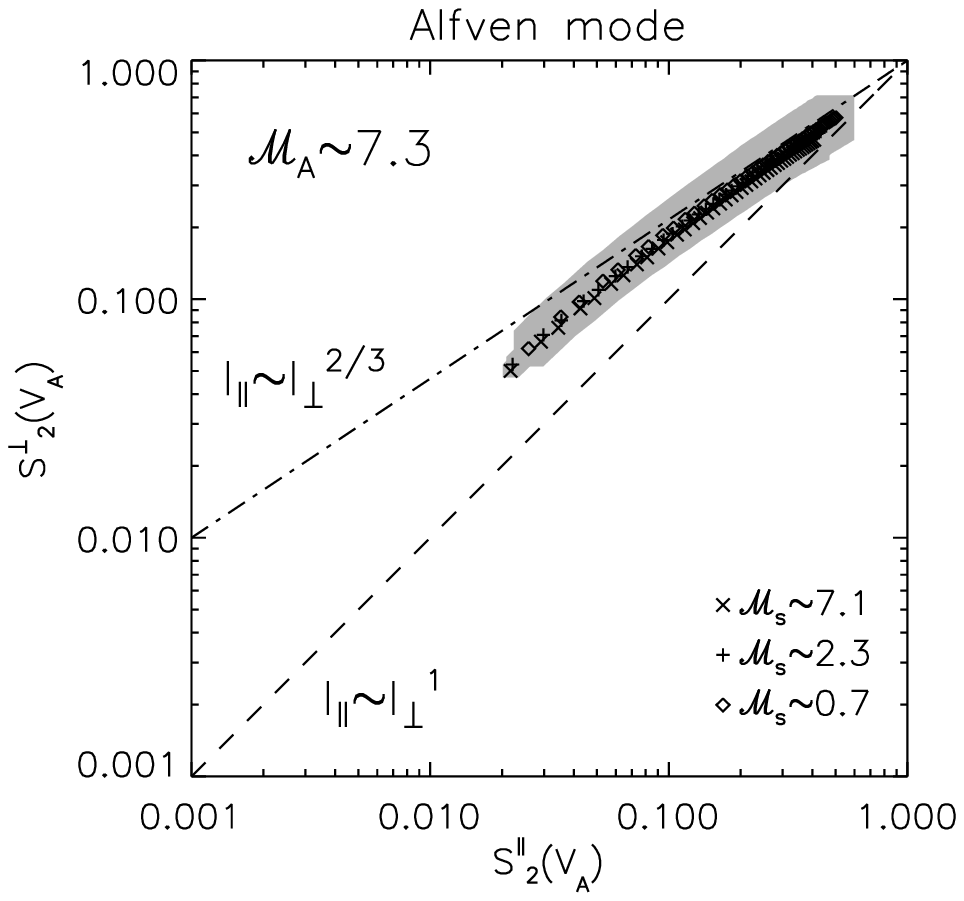}
 \plotone{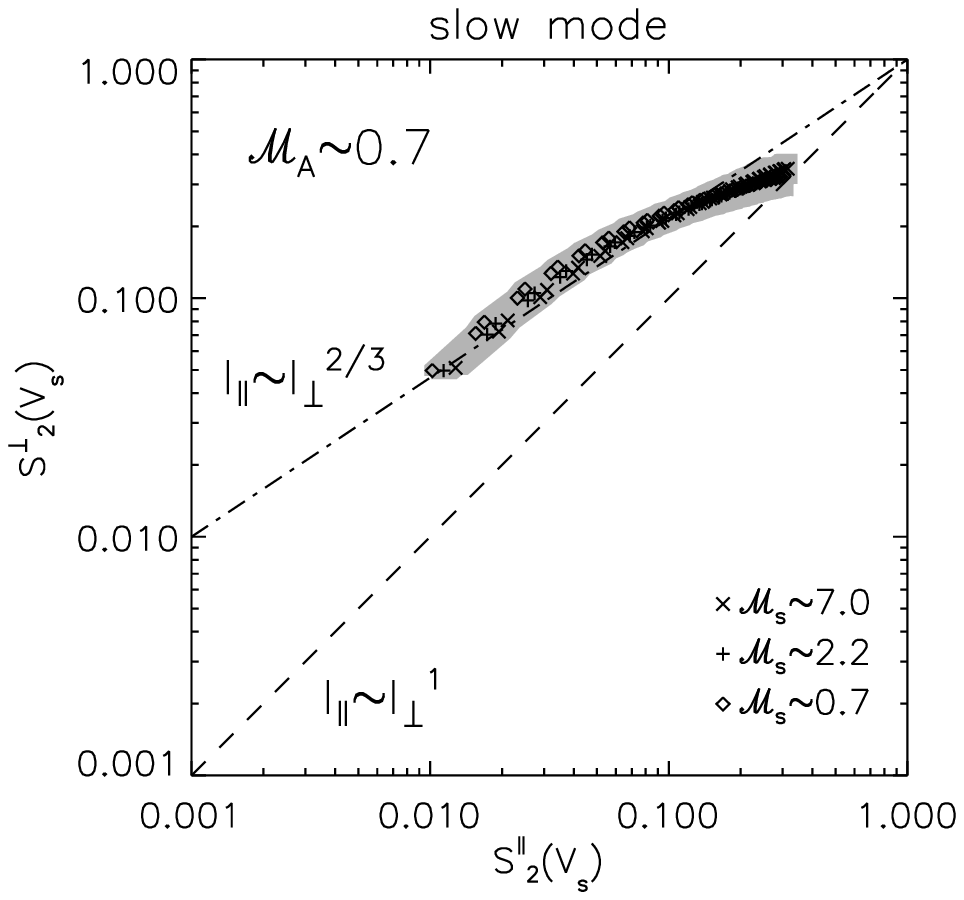}
 \plotone{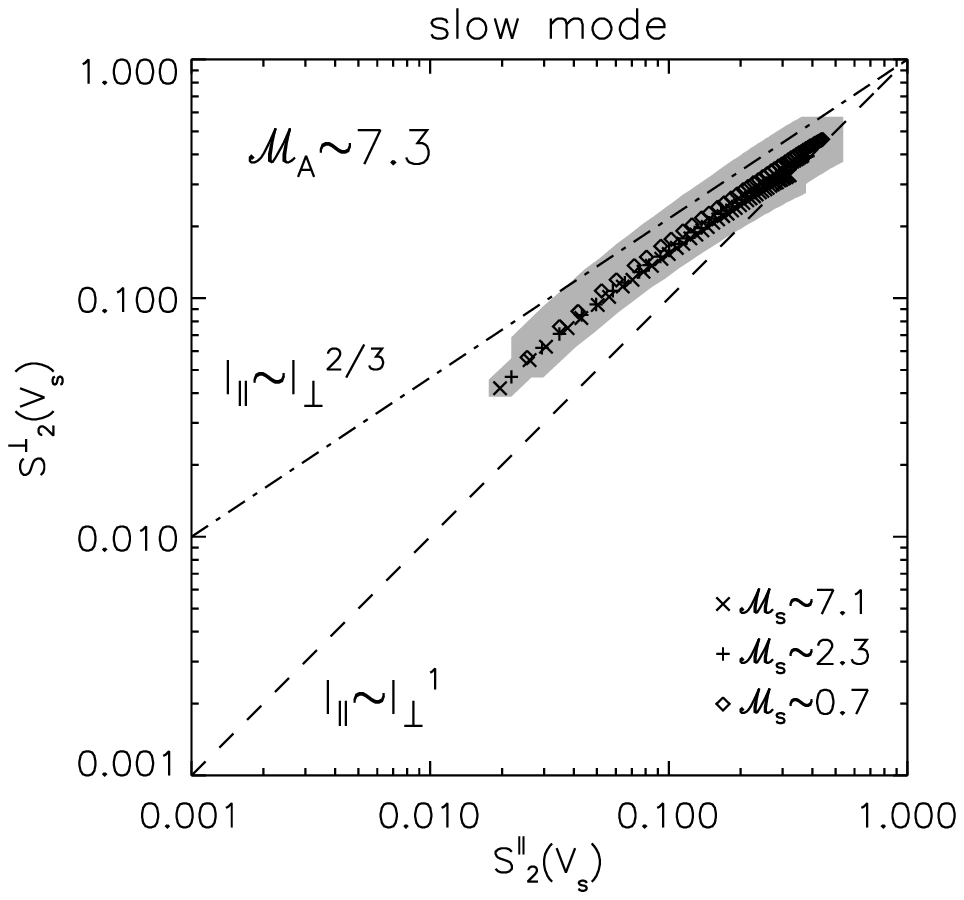}
 \plotone{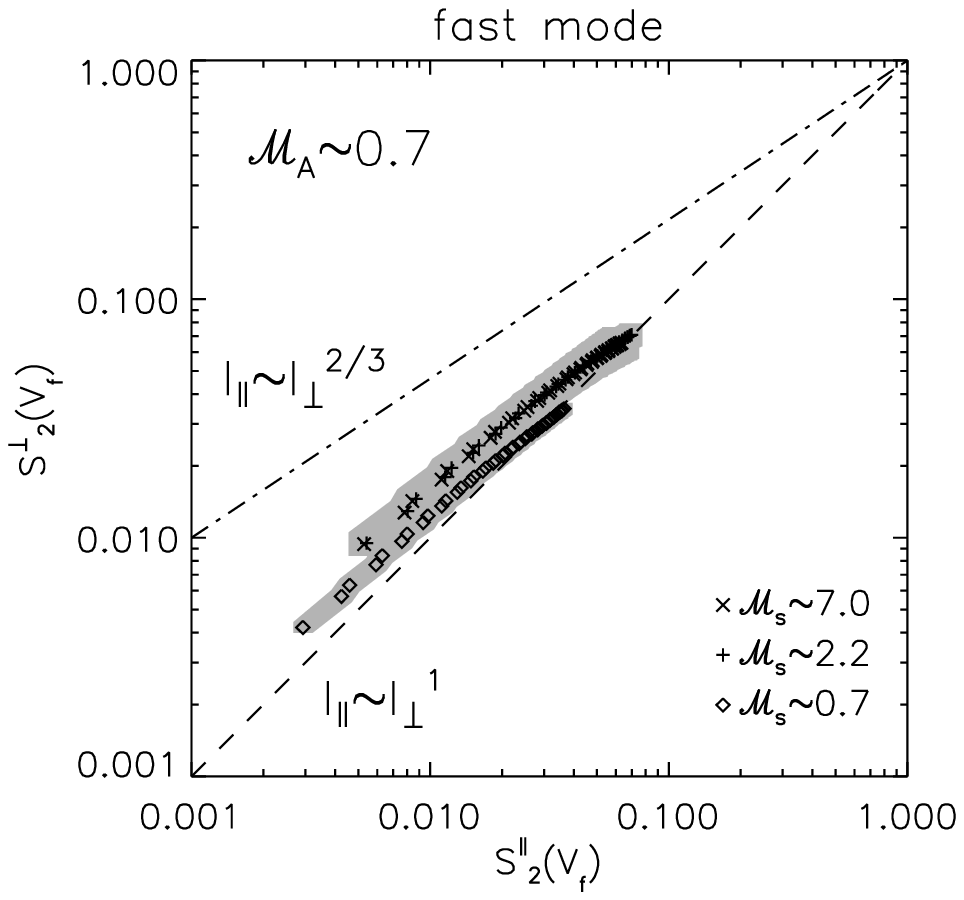}
 \plotone{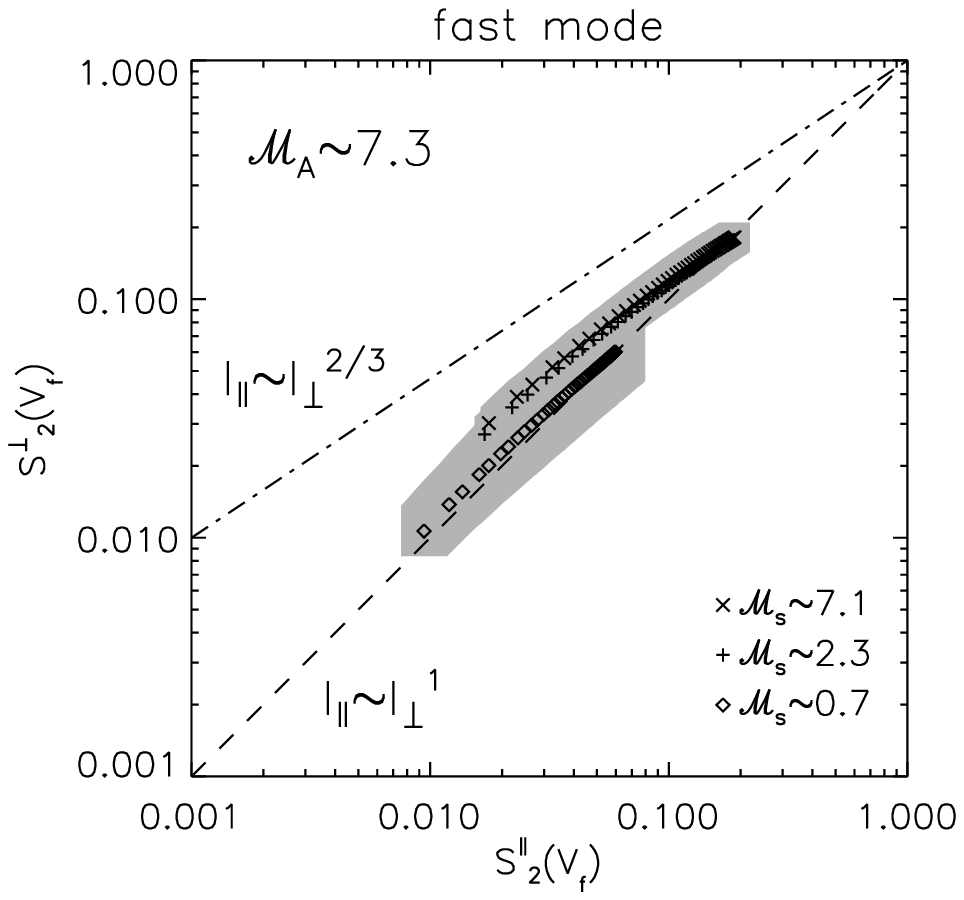}
 \caption{Anisotropy of the structures of the Alfv\'{e}n, slow and fast modes (top, middle and bottom rows, respectively). To show the anisotropy we use the 2$^\mathrm{nd}$-order total structure functions, parallel and perpendicular to the local mean magnetic field. Points correspond to the mean profiles of the structure functions averaged over several snapshots. The gray areas under points correspond to the degree of departures of the structure functions in time. \label{fig:mode_anisotropy}}
\end{figure*}

In Figure \ref{fig:comp_anisotropy} we present the degree of anisotropy for the
velocity field (top row) and for the solenoidal (incompressible) and potential
(compressible) parts of the velocity fields (middle and bottom row,
respectively).  The left and right columns correspond to the sub- and
superAlfv\'{e}nic turbulence, respectively.  The gray areas under points show
the degree of time variances of the structure functions used to plot
anisotropies.  In each plot of Figure~\ref{fig:comp_anisotropy} we also plot
lines corresponding to the isotropic structure, $l_\parallel \sim l_\perp$, and
theoretical, $l_\parallel \sim l_\perp^{2/3}$ \citep[][GS95]{goldreich95}.  We
see that the structures of velocity and its incompressible part show the GS95
type anisotropy, for both Alfv\'{e}nic regimes. Another conclusion coming from
these plots is that the anisotropy is insensitive on the value of the sonic Mach
number.  All curves have the same shape in the presented plots. The one
noticeable difference is that in the case of subAlfv\'{e}nic turbulence the
anisotropy slightly changes with the values of $S^{\parallel}_2$, which are
related to the scale of structures.  Lower values of $S^{\parallel}_2$
correspond to the small-scale structures, larger to the large-scale structures.
In this description we can suspect, that the small scale structures are almost
isotropic.  The degree of anisotropy grows with the scale, and for the large
scale structures it exceeds somewhat the GS95 anisotropy.  In the case of a weak
external magnetic field the degree of anisotropy is uniform over the whole range
of scales, but also here we notice very good agreement with GS95 anisotropy.
The compressible component of the velocity field behaves differently.  For
subAfv\'{e}nic turbulence it is almost perfectly isotropic and independent of
the sonic Mach number.  In the superAlfv\'{e}nic turbulence, however, we see,
that structure of subsonic potential field is isotropic on average, although
there are relatively large time departures from the isotropy.  On contrary, in
the case of supersonic turbulence, the structure of the potential field contains
a higher amount of anisotropic structures and the degree of anisotropy aims to
the GS95-law.

In Figure \ref{fig:mode_anisotropy} we show the anisotropy of structures for
Alfv\'{e}n, slow and fast modes of velocity.  The separation of the modes is
performed with the respect of the local mean magnetic field.  The first
comparison of these plots shows, that the degree of anisotropy does not depend
on the sonic Mach number for the Alfv\'{e}n and slow modes.  The anisotropy of
fast mode, similarly to the case of the compressible part of the velocity,
depends on the regime of sonic motions.  If the fluid motions are subsonic the
structures of the fast waves is more isotropic.  When the fluid motions are
supersonic, more structures become anisotropic.  We should note here, that
speaking about the anisotropy of structures we understand a mean anisotropy of
all structures of the analyzed field.  Individual structures in the turbulence
evolve and yield at different deformations resulting in a constant change of the
degree of the anisotropy.  In this situation, speaking about an anisotropy, we
are relating to the mean dominant anisotropy of all structures in the system.

The two remaining modes, the Alfv\'{e}n and slow, show very similar anisotropies
to those observed in velocity and its incompressible part.  We see that
basically the degree of anisotropy of structures in these two modes is closer to
GS95.  In addition, the presence of a strong magnetic field results in a much
stronger bending of the curves in right plots of
Figure~\ref{fig:mode_anisotropy} signifying changes of the anisotropy with the
scale of the structure.  The curves in plots for superAlfv\'{e}nic models are
almost straight and independent of the values of $S^{\parallel}_2$.

\section{Scaling Exponents and Intermittency}
\label{sec:intermittency}

Intermittency is an essential property of astrophysical fluids.  As
intermittency violates self-similarity of motions, it is impossible to naively
extrapolate the properties of fluids obtained computationally with a relatively
low resolution to the actual astrophysical situations.  In astrophysics the
intermittency affects turbulent heating, momentum transfer, interaction with
cosmic rays, radio waves and many more essential processes.  Physical
interpretation of intermittency started after the work by Kolmogorov, but the
first successful model was presented by \cite{she94}.  The scaling relations
suggested by \cite{she94} relate $\zeta(p)$ to the scaling of the velocity $v_l
\sim l^{1/g}$, the energy cascade rate $t^{-1}_l \sim l^{-x}$, and the
co-dimension of the dissipative structures $C$:
\begin{equation}
 \zeta(p) = \frac{p}{g} (1-x) + C \left( 1 - (1 - x/C)^{p/g} \right).
\end{equation}
Parameter $C$ is related to the dimension of the dissipative structures $D$
through relation $C = 3 - D$ \citep{mueller00}.  In hydrodynamical turbulence,
according to the Kolmogorov scaling, we have $g = 3$ and $x = 2/3$.  Vortex
filaments, which are one-dimensional structures, correspond to $C=2$ ($D=1$).
In the MHD turbulence we also observe current sheets, which are two-dimensional
dissipative structures and correspond to $C=1$ ($D=2$).  For these two types of
dissipative structures we obtain two different scaling relations (substituting
$g = 3$ and $x = 2/3$):
\begin{equation}
 \zeta(p) = \frac{p}{9} + 2 \left[ 1 - (2/3)^{p/3} \right] \ \mathrm{for} \ C=2
 \label{eqn:sl}
\end{equation}
and
\begin{equation}
 \zeta(p) = \frac{p}{9} + 1 - (1/3)^{p/3} \ \mathrm{for} \ C=1 .
 \label{eqn:mb}
\end{equation}
Relation (\ref{eqn:sl}) is often called the She \& L\'{e}v\^{e}que scaling
\citep{she94}, while relation (\ref{eqn:mb}), the M\"uller-Biskamp scaling
\citep{mueller00}.  There are theoretical arguments against the model
\citep[see][]{novikov94}, but so far the She \& L\'{e}v\^{e}que scaling is the
best for reproducing the intermittency of incompressible hydrodynamic
turbulence.

Structure functions can be calculated with respect to the global or local
reference frames.  By the scaling exponents calculated in the global reference
frame we understand the scaling exponents calculated from the structure
functions averaged over all directions.  In the local reference we distinguish
between directions parallel and perpendicular to the local mean magnetic field.
In this way we define the reference frame locally by the direction of the local
magnetic field.
\begin{figure*}  
 \epsscale{0.36}
 \plotone{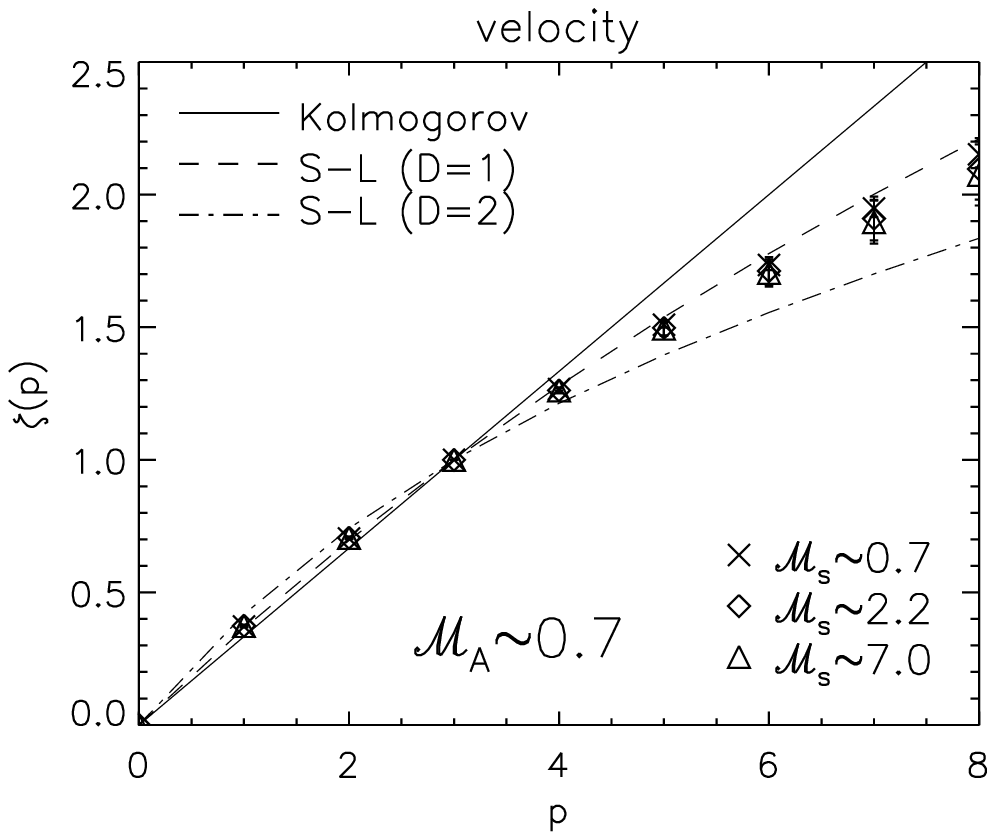}
 \plotone{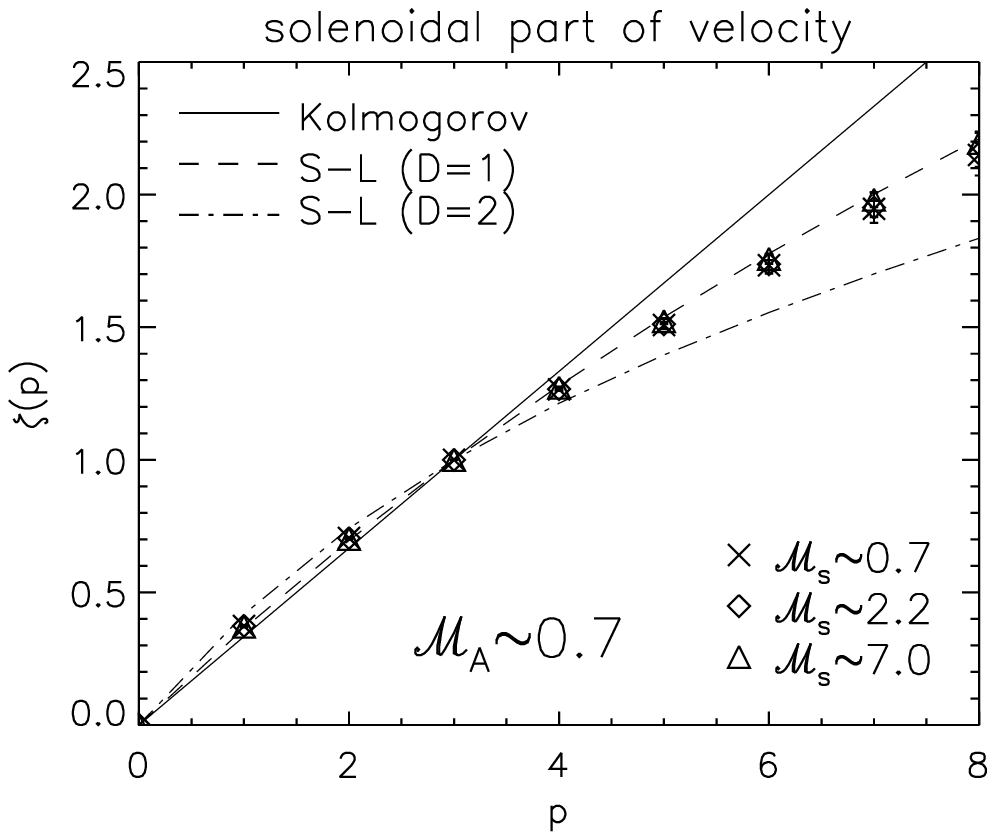}
 \plotone{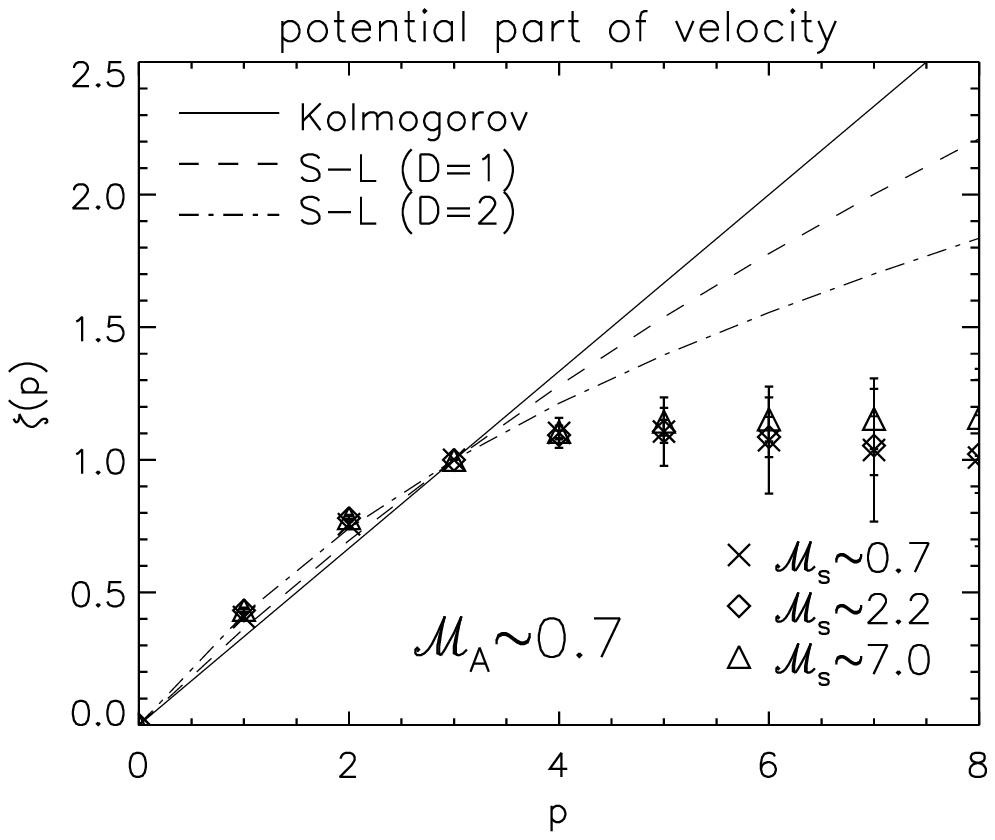}
 \plotone{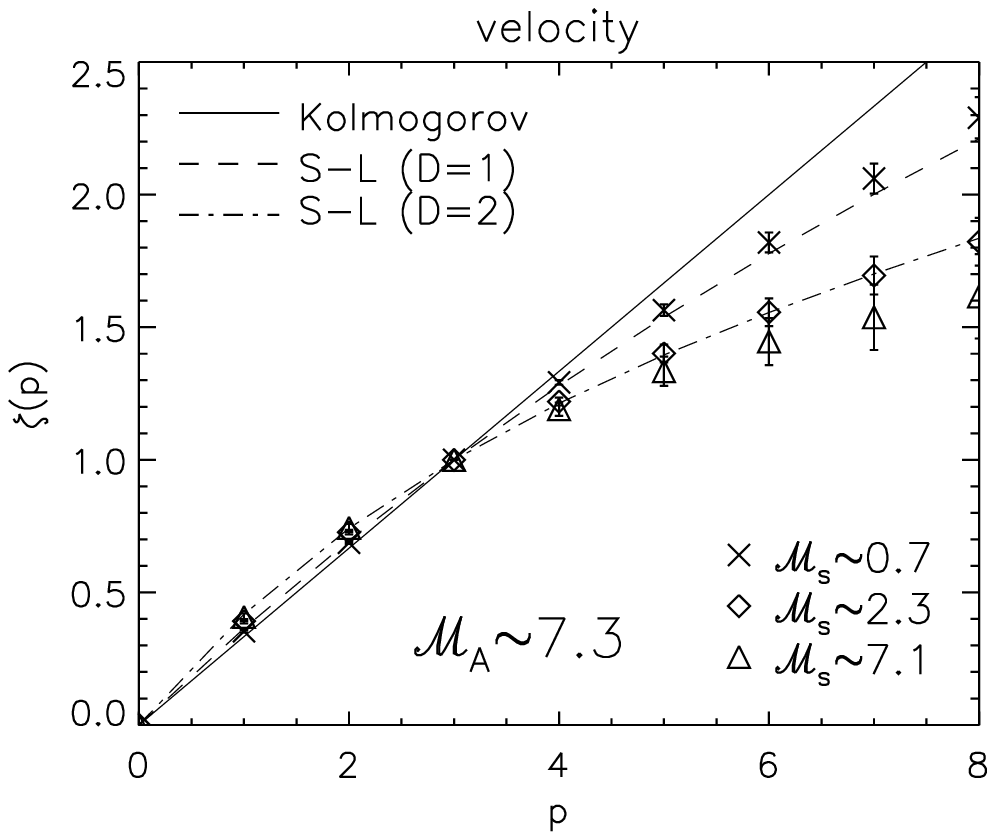}
 \plotone{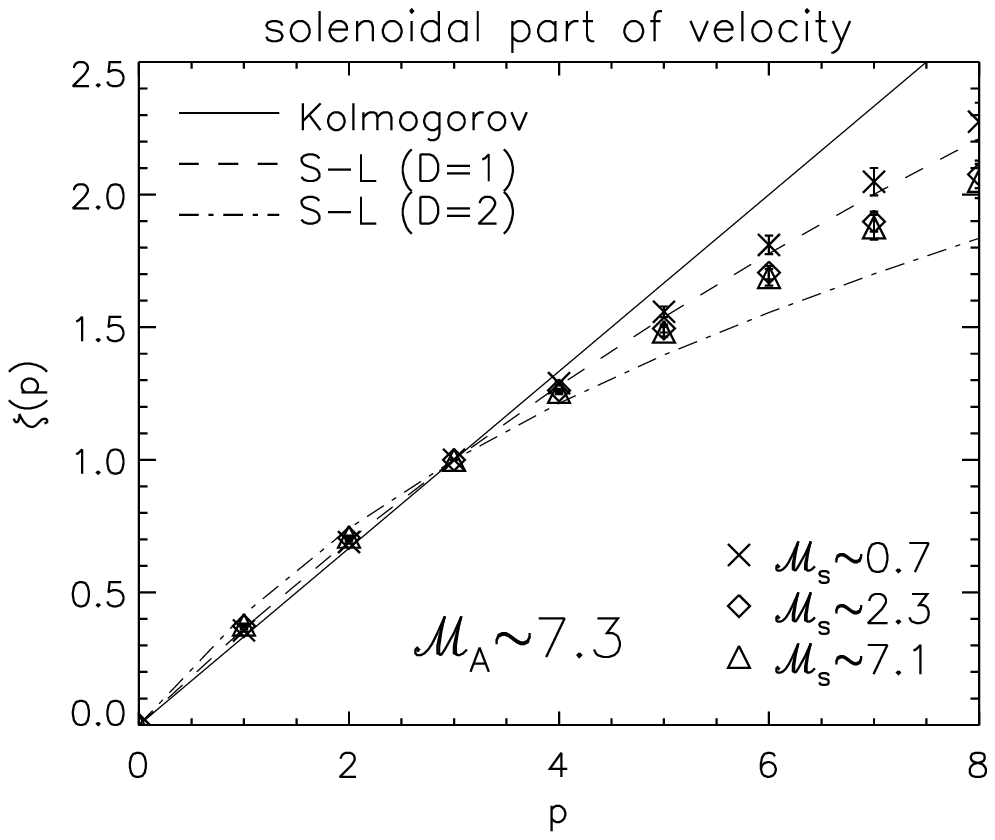}
 \plotone{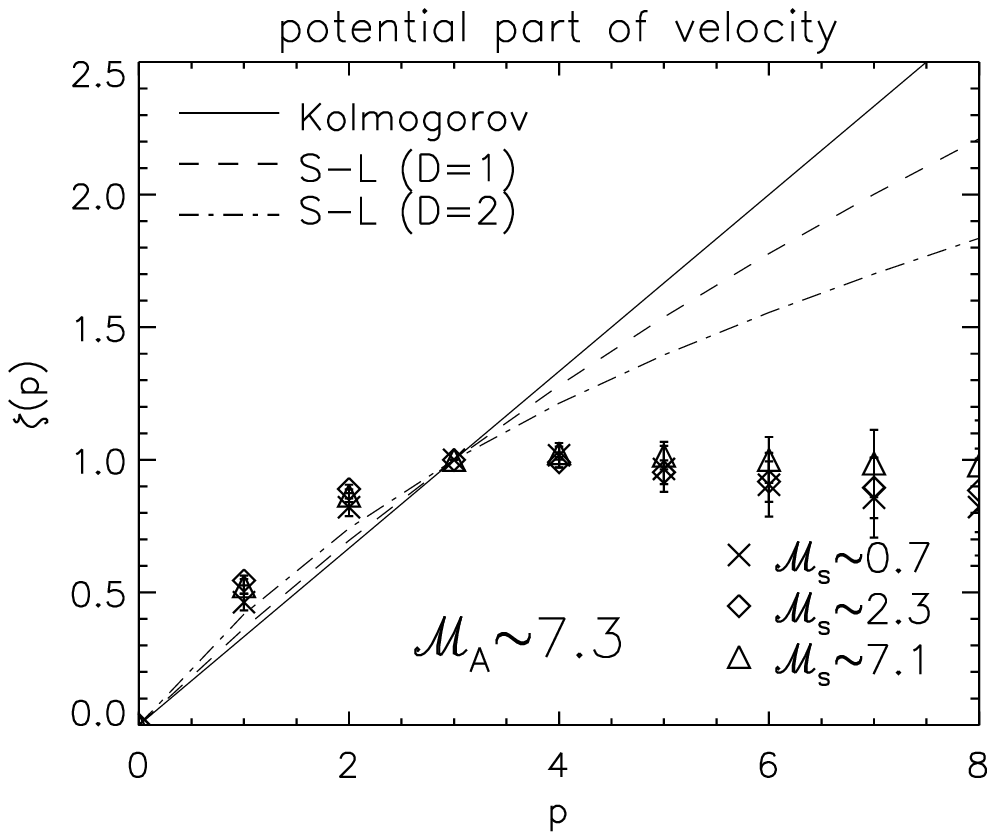}
 \caption{Scaling exponents of the velocity (left column) and its incompressible and compressible parts (middle and right columns, respectively) for experiments with different sonic Mach numbers in two regimes: subAlfv\'{e}nic (upper row) and superAlfv\'{e}nic (lower row). \label{fig:expons_parts}}
\end{figure*}

\begin{figure*}  
 \epsscale{0.36}
 \plotone{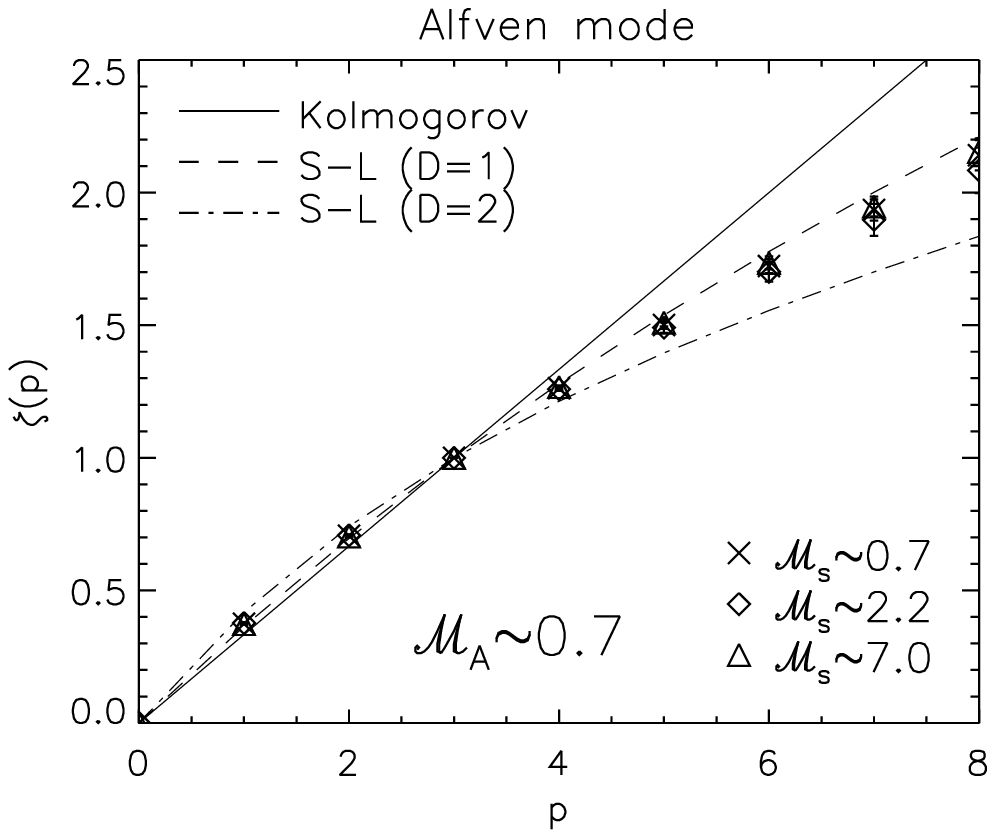}
 \plotone{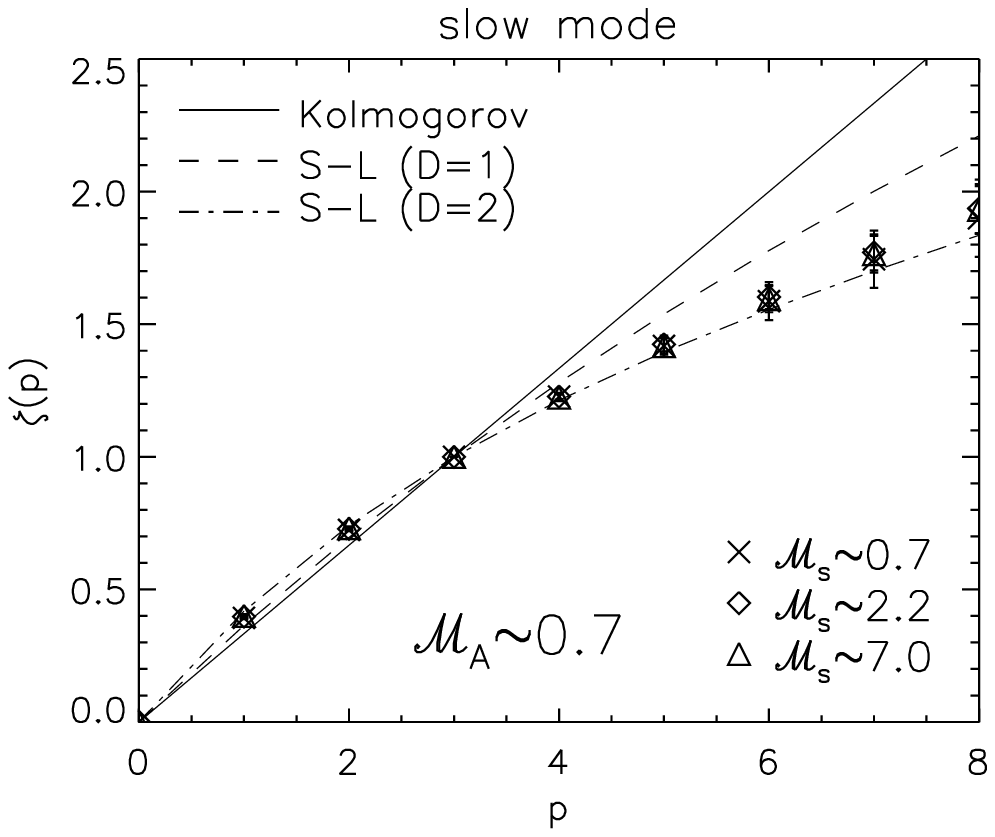}
 \plotone{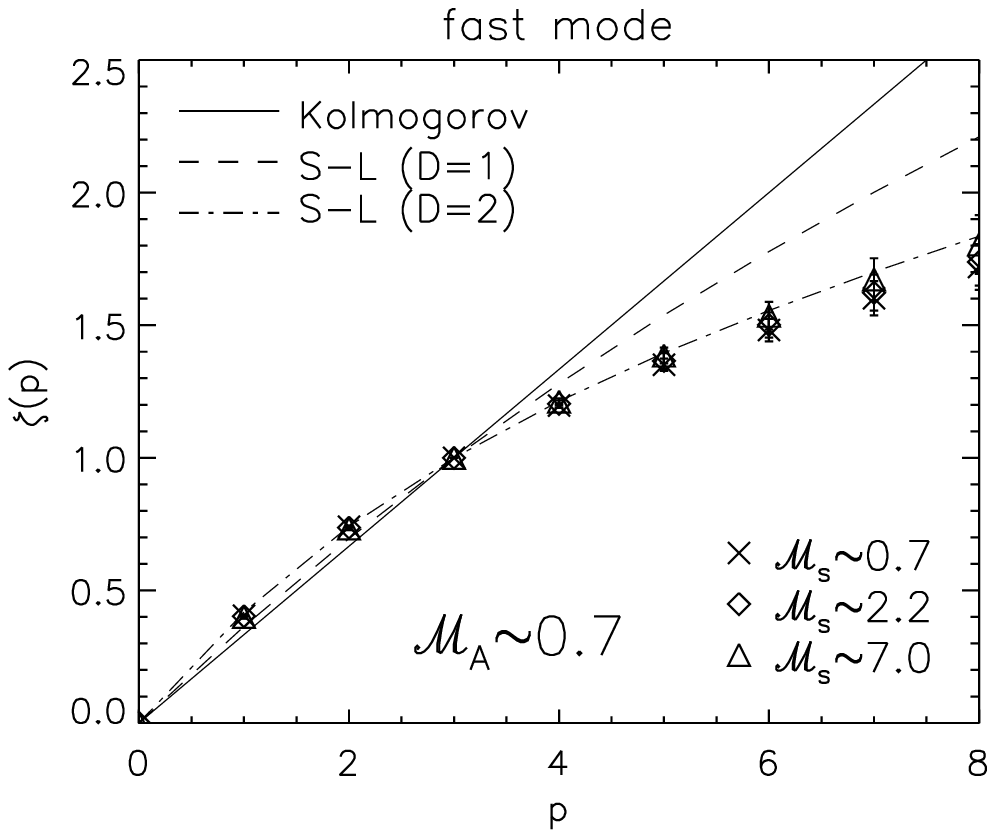}
 \plotone{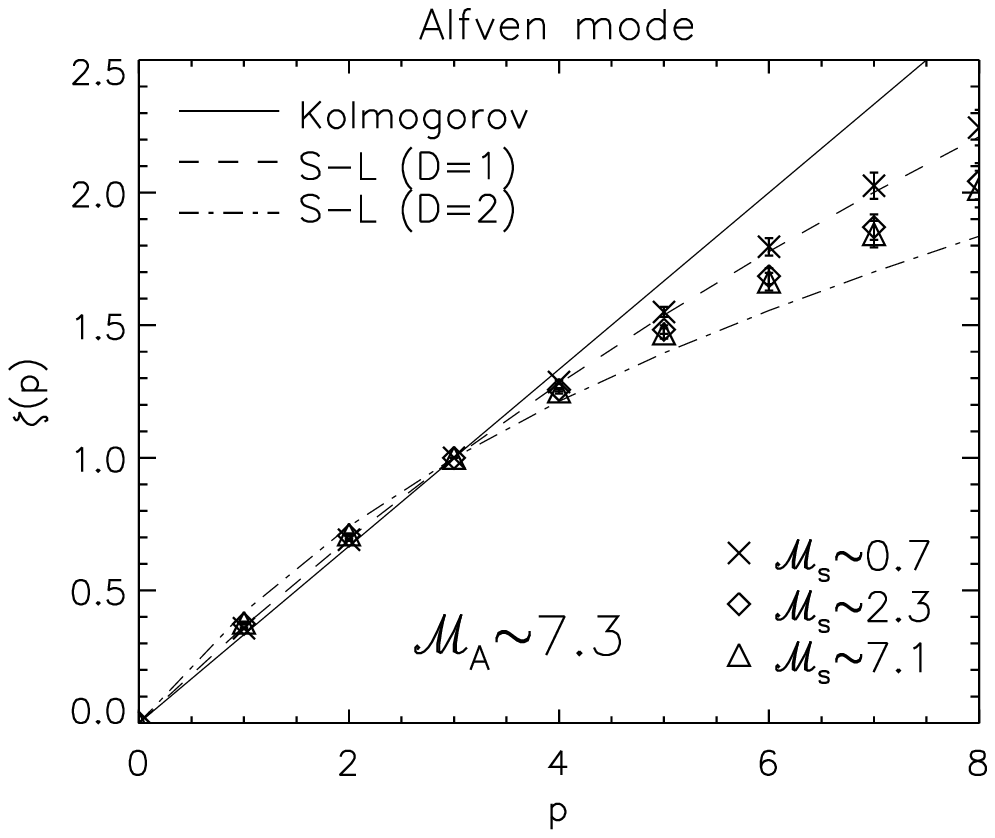}
 \plotone{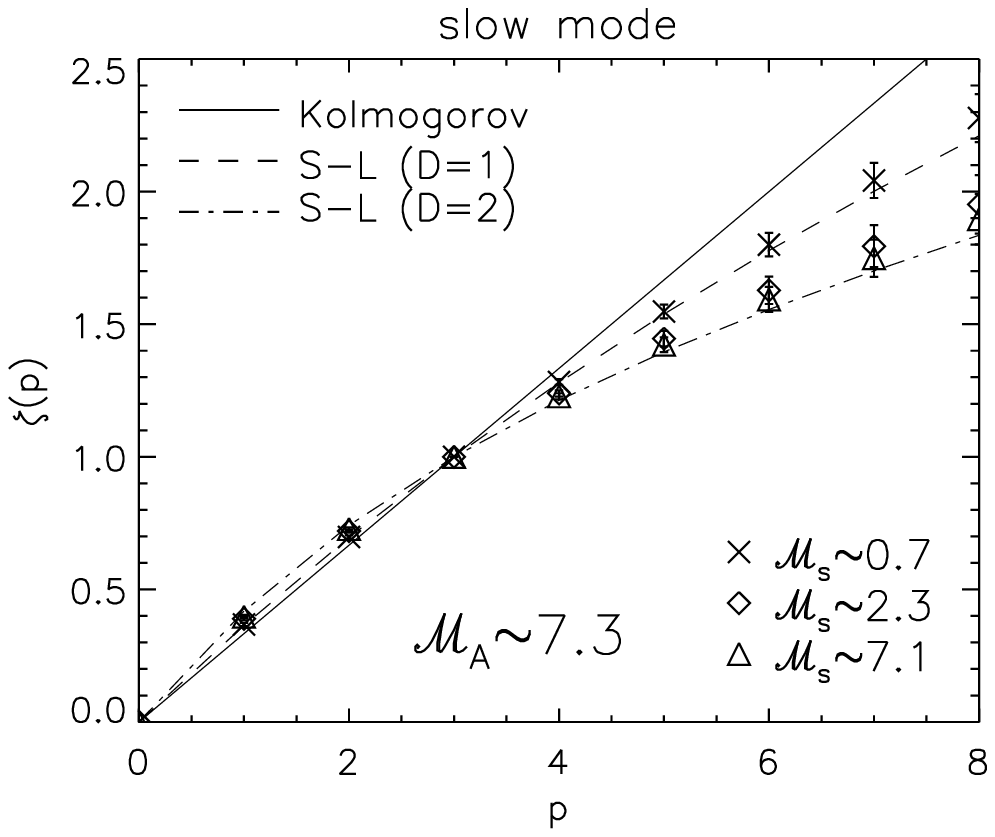}
 \plotone{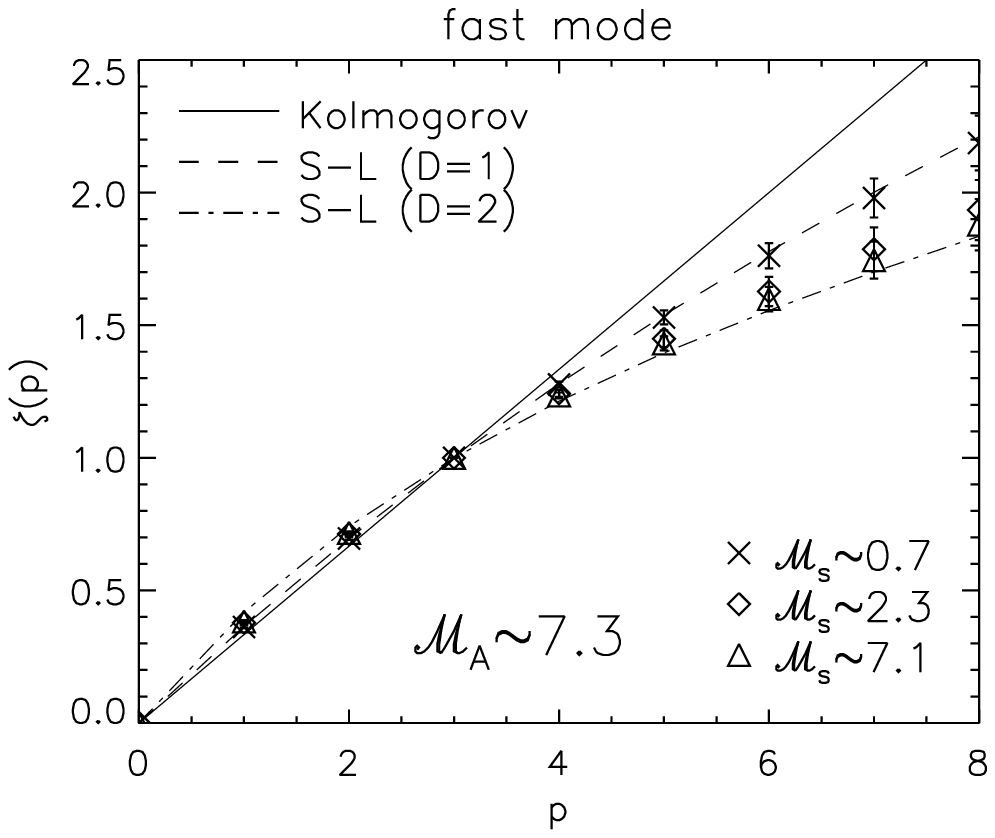}
 \caption{Scaling exponents of the Alfv\'{e}n, slow and fast modes of the velocity (left, middle and right columns, respectively) for experiments with different sonic Mach numbers in two regimes: subAlfv\'{e}nic (upper row) and superAlfv\'{e}nic (lower row). \label{fig:expons_waves}}
\end{figure*}

In Figures~\ref{fig:expons_parts} and \ref{fig:expons_waves} we show scaling
exponents for the velocity and all its parts and waves calculated in the global
reference frame.  In the top left plot of Figure~\ref{fig:expons_parts} we see
that for the subAlfv\'{e}nic turbulence the scaling exponents of velocity follow
the She-L\'{e}v\^{e}que (S-L) scaling with $D=1$.  Supported by the theoretical
considerations we can say that most of the dissipative structures are
one-dimensional.  Even though the scalings are not perfectly independent of the
value of ${\cal M}_s$, since we see somewhat lower values of $\zeta$ for higher
$p$, the differences between these values for models with different sonic Mach
numbers are within their error bars, thus it is relatively difficult to state
that the scalings are completely independent or only weakly dependent of the
values of ${\cal M}_s$.  Looking in the corresponding plot for models with a
weak magnetic field we clearly see that the spread of curves for different sonic
Mach numbers is much higher than in the previous case.  For subsonic model the
scaling exponents of velocity follow very well the theoretical curve defined by
the S-L scaling with parameter $D$ corresponding to one-dimensional structures.
The model with ${\cal M}_s \sim 2.3$, however, follows perfectly the S-L scaling
with $D=2$ corresponding to the two-dimensional dissipative structures.
Moreover, models with even higher values of the sonic Mach number have the
scaling exponents for $p>3$ somewhat below the S-L scaling with $D=2$.  These
observations suggest that the scaling exponents of the velocity change with the
sonic Mach number but only in the case of weak magnetic field turbulence.  The
presence of a strong magnetic field significantly reduces these changes and
preserves the generation of the dissipative structures of higher than one
dimensions.

After the decomposition of velocity into its incompressible and compressible
parts we also calculate their scaling exponents.  In the middle and right
columns of Figure~\ref{fig:expons_parts} we show the incompressible and
compressible parts of the velocity field, respectively.  The incompressible part
it strong.  It constitutes most of the velocity field thus it is not surprising
that its scaling exponents are very similar to those observed in velocity.  This
is true in the case of subAlfv\'{e}nic models, because all curves in the middle
left plot in Figure~\ref{fig:expons_parts} are tightly covering the S-L scaling
with $D=1$.  The similarity between the velocity and its solenoidal part is also
confirmed in the case of superAlfv\'{e}nic models but only for subsonic case,
when the role of shocks is strongly diminished.  Two supersonic models show
exponents following a scaling more closer to the S-L one with $D=1$, yet still
with lower values for $p>3$.  The scalings of structure of the compressible
part, shown in the right column of Figure~\ref{fig:expons_parts}, cannot be
compared to any of the theoretical models.  Their scalings signify dissipative
structures with dimensions higher than two, but the theory may be not applicable
here.  In order to explain what these scaling exponents represent we should
describe what could be a physical picture of the compressible part of the
velocity field.  First of all, the compressible part is much weaker then the
incompressible one, thus the structure functions of higher orders amplify rare
events in the structure, such as individual regions compressed by shocks.  This
is supported by the huge error bars increasing with the value of $p$ signifying
that these rare events may have poor statistics or can change rapidly
contributing differently at different times and in different models.  In such
situations the scaling exponents for higher values of $p$ are not reliable.

Another decomposition, which separate velocity into the MHD waves, gives an
opportunity to calculate their scaling relations as well.  In
Figure~\ref{fig:expons_waves} we present the scaling exponents for the
Alfv\'{e}n, slow and fast waves.  The Alfv\'{e}n wave is presented in plots in
the left column.  Comparing these plots with the corresponding plots for the
solenoidal part of the velocity reveals obvious conclusion that the Alfve\'{e}n
wave due to its incompressibility should follow exactly the same scaling as the
incompressible part.  This is visible in these two plots clearly.  Scaling
exponents of the solenoidal part and the Alfv\'{e}n mode match nicely in plots
for subAlfv\'{e}nic models with a weaker dependency of the sonic Mach number and
for superAlfv\'{e}nic models where the dependency of ${\cal M}_s$ is stronger.
Two other modes, the slow and fast waves presented in the middle and right
columns of Figure~\ref{fig:expons_waves}, respectively, show similar scaling
relations.  For example, the scaling exponents for slow and fast waves in the
subAlfv\'{e}nic turbulence follow the S-L scaling with $D=2$ and depend on the
sonic Mach number marginally.  This signifies that the two-dimensional
structures dominate in both components.  For superAlfv\'{e}nic turbulence the
scalings of these two waves show a somewhat different situation.  Both
components, the slow and fast waves, have similar values of scaling exponents
$\zeta$, but they change with the sonic Mach number.  We see that for subsonic
turbulence most of dissipative structures of the slow and fast waves is
one-dimensional.  Scaling exponents follow very precisely the S-L scaling with
$D=1$.  In the case of supersonic turbulence, however, these scalings suggest
the two-dimensional dissipative structures again, similarly to the
subAlfv\'{e}nic turbulence.  This signifies an important role of the magnetic
field in the generation of the structures of different dimensions in the
subsonic turbulence.

\begin{figure*}  
 \epsscale{0.36}
 \plotone{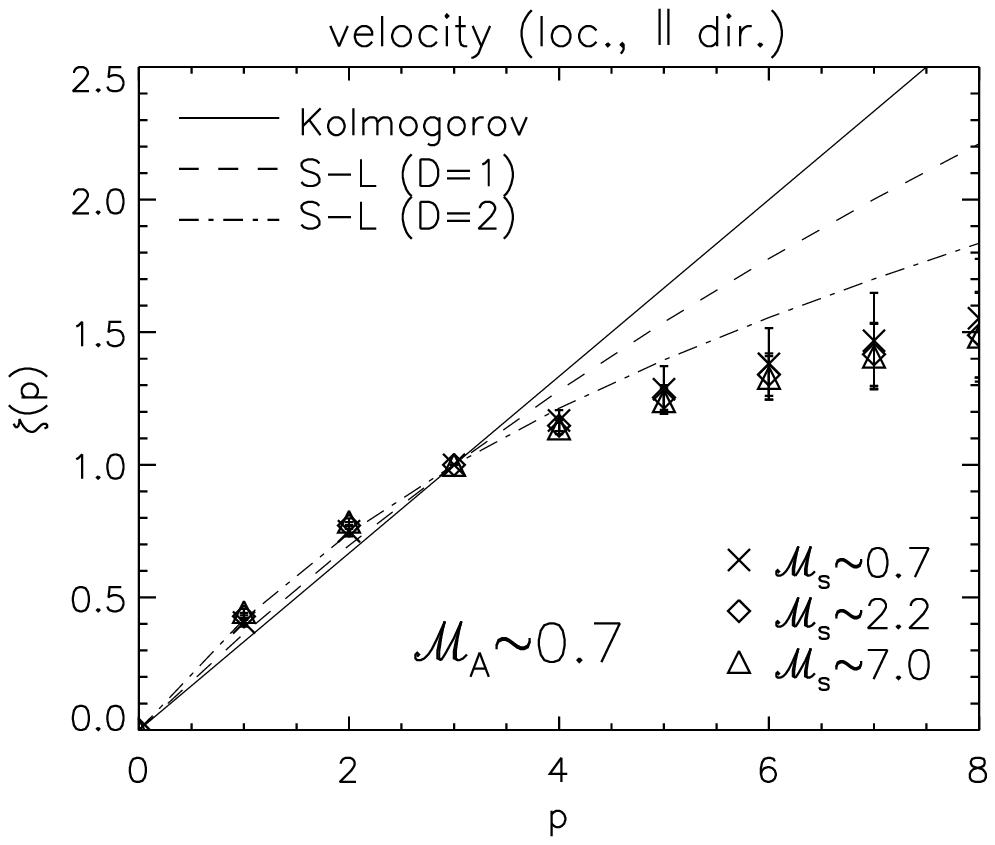}
 \plotone{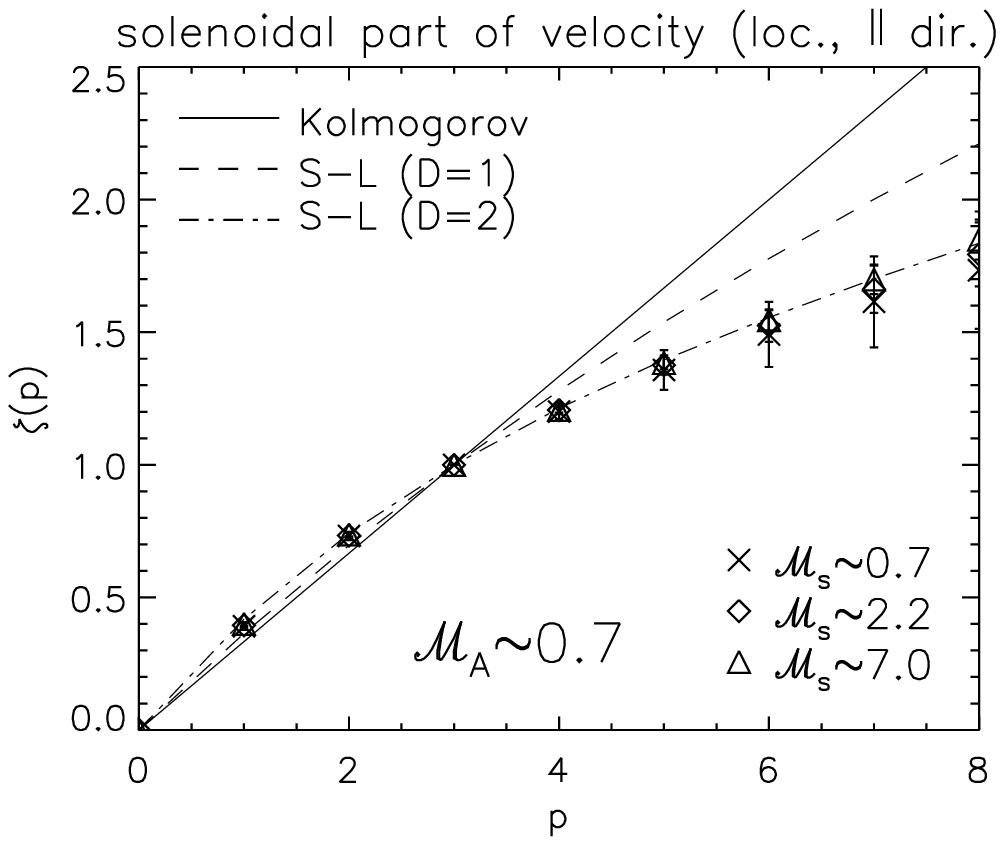}
 \plotone{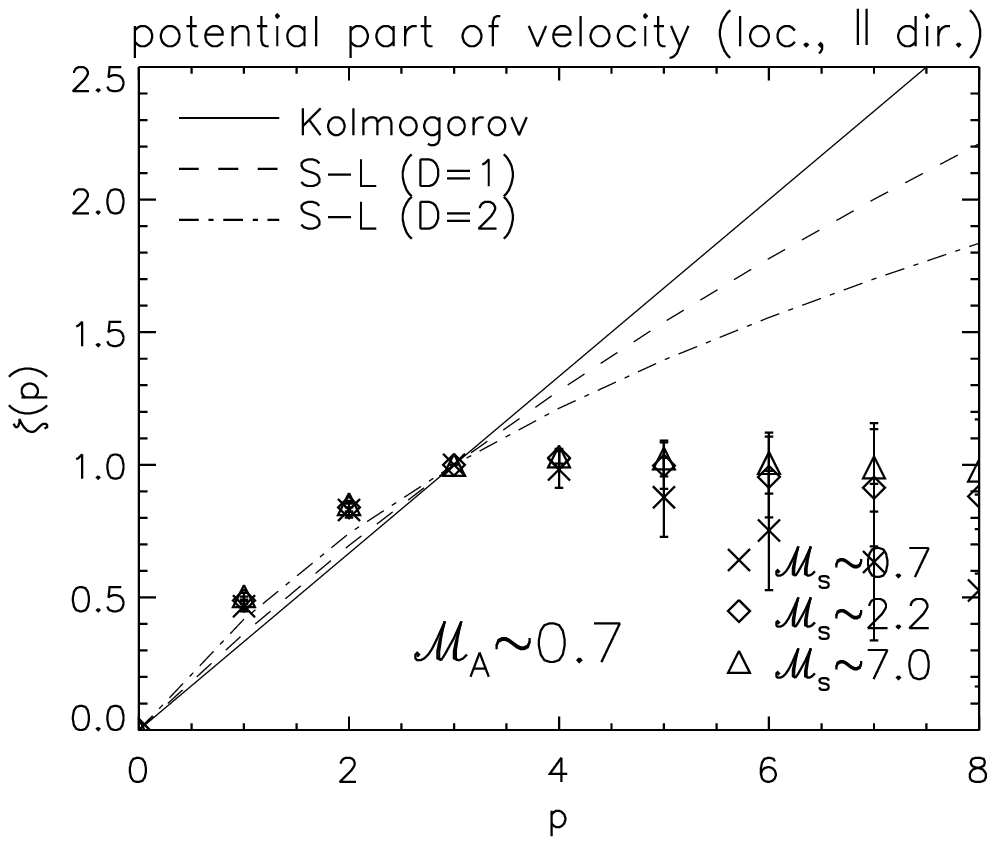}
 \plotone{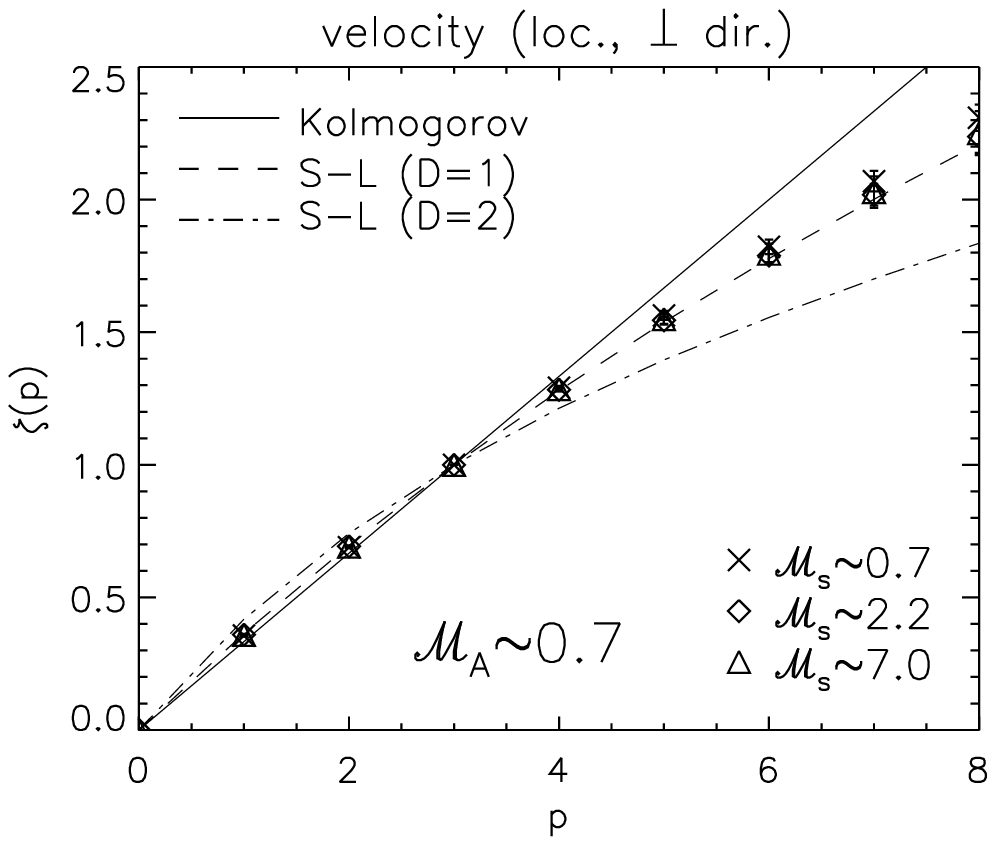}
 \plotone{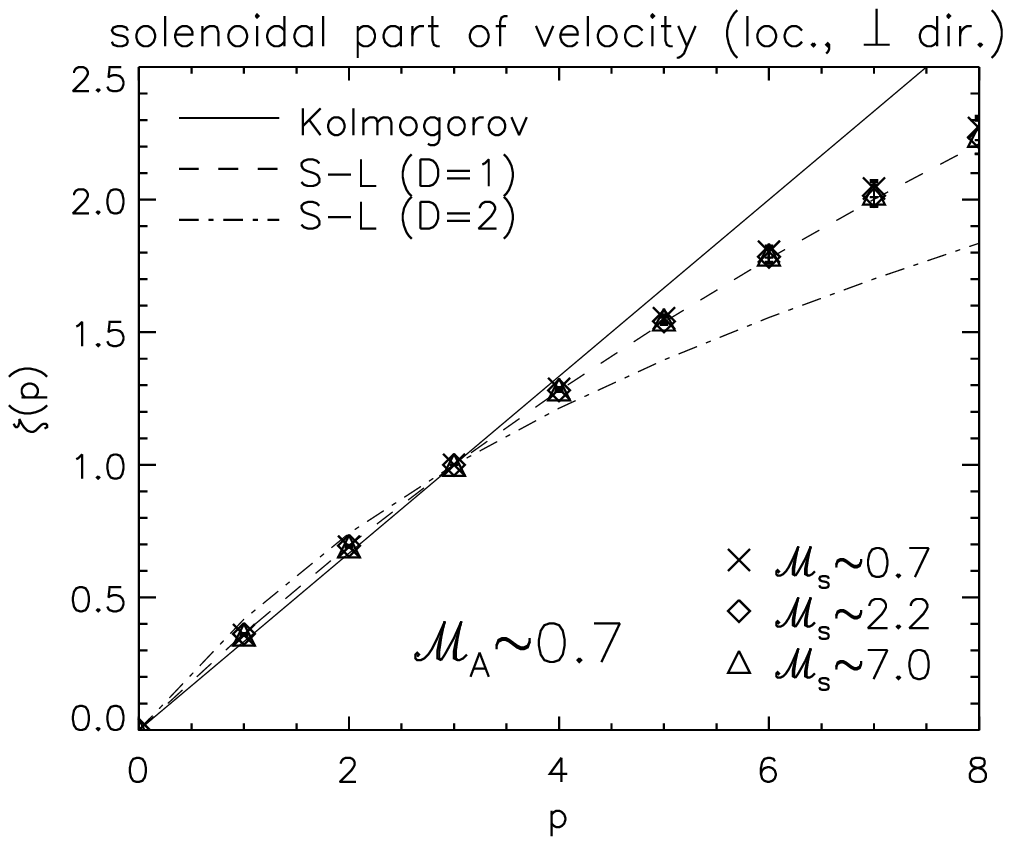}
 \plotone{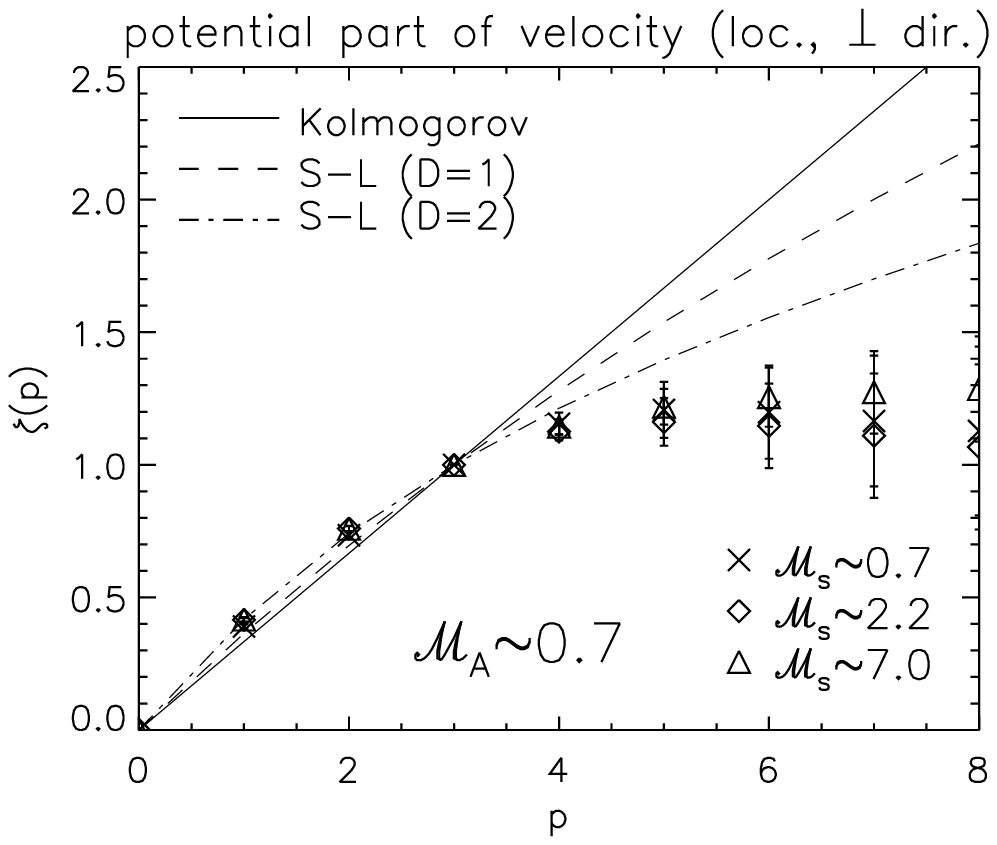}
 \caption{Scaling exponents of the velocity and its solenoidal and potential parts (left, middle and right columns, respectively) calculated in the local reference frame.  Two rows show the scalings exponent at the parallel and perpendicular direction the local mean magnetic field for models with a strong magnetic field. \label{fig:expons_parts_local}}
\end{figure*}

\begin{figure*}  
 \epsscale{0.36}
 \plotone{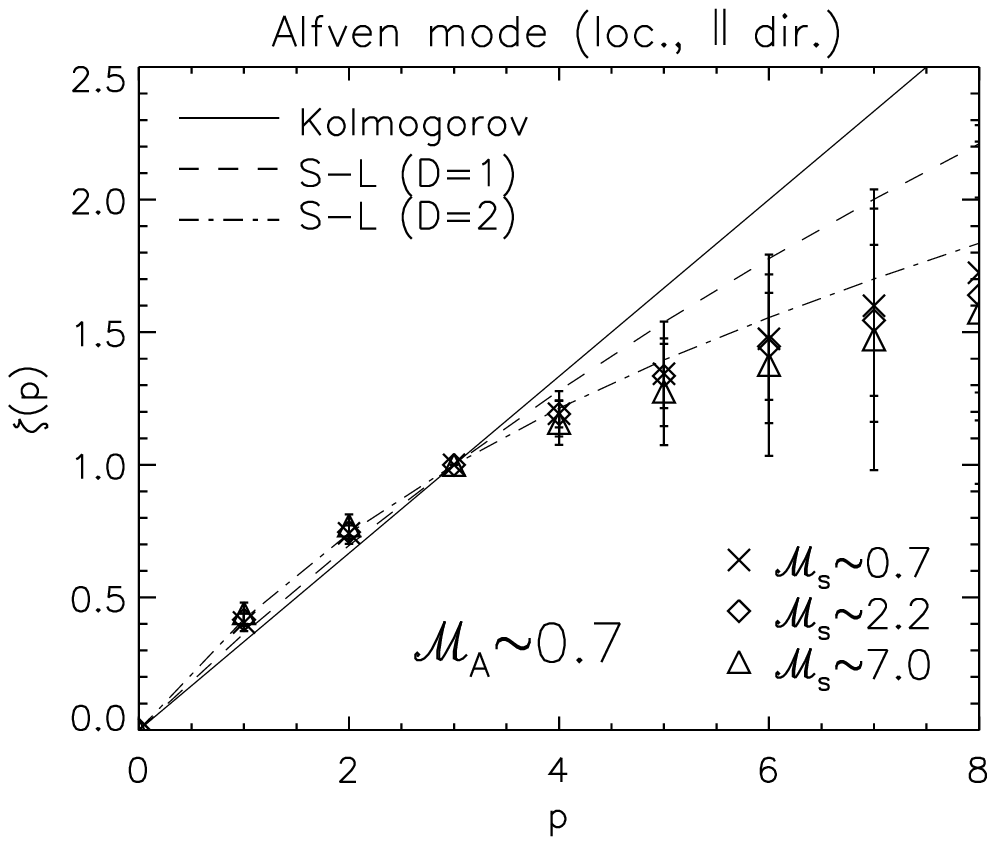}
 \plotone{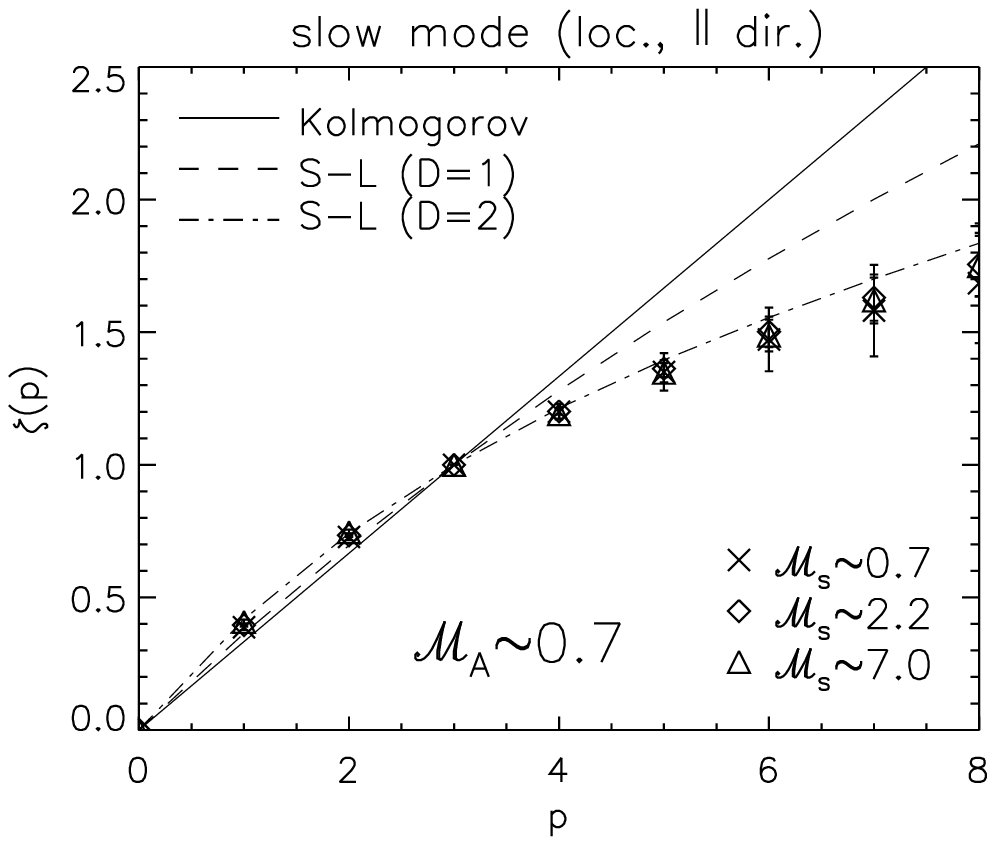}
 \plotone{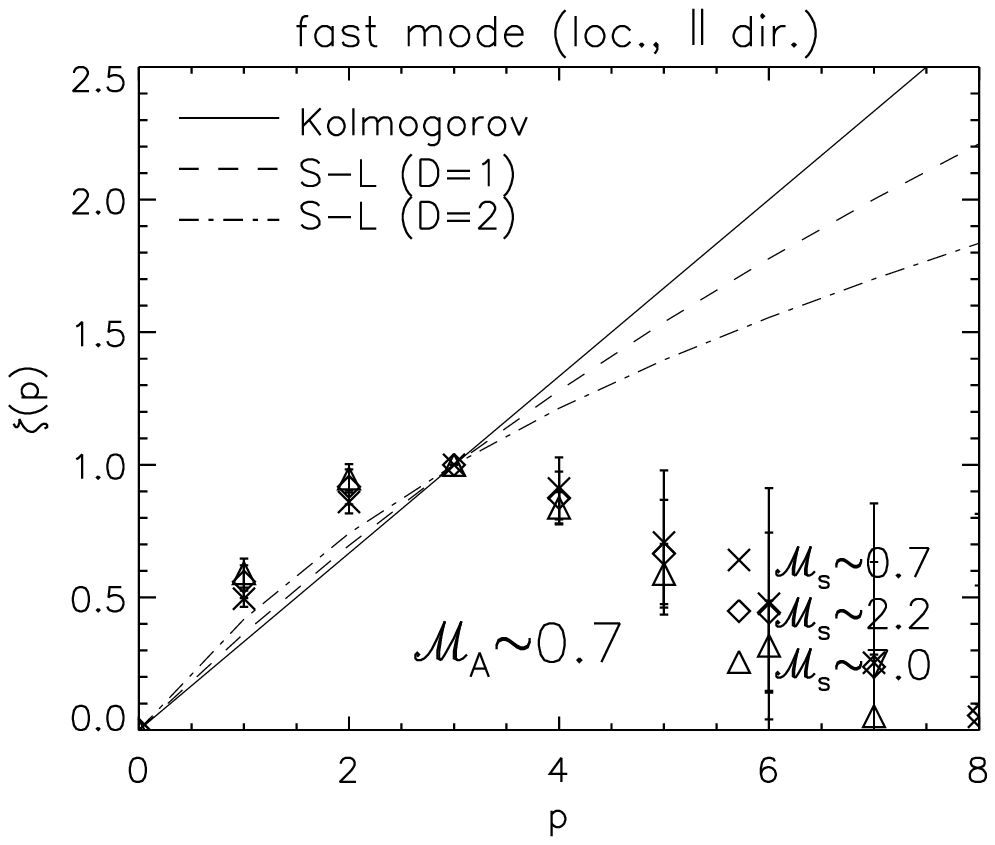}
 \plotone{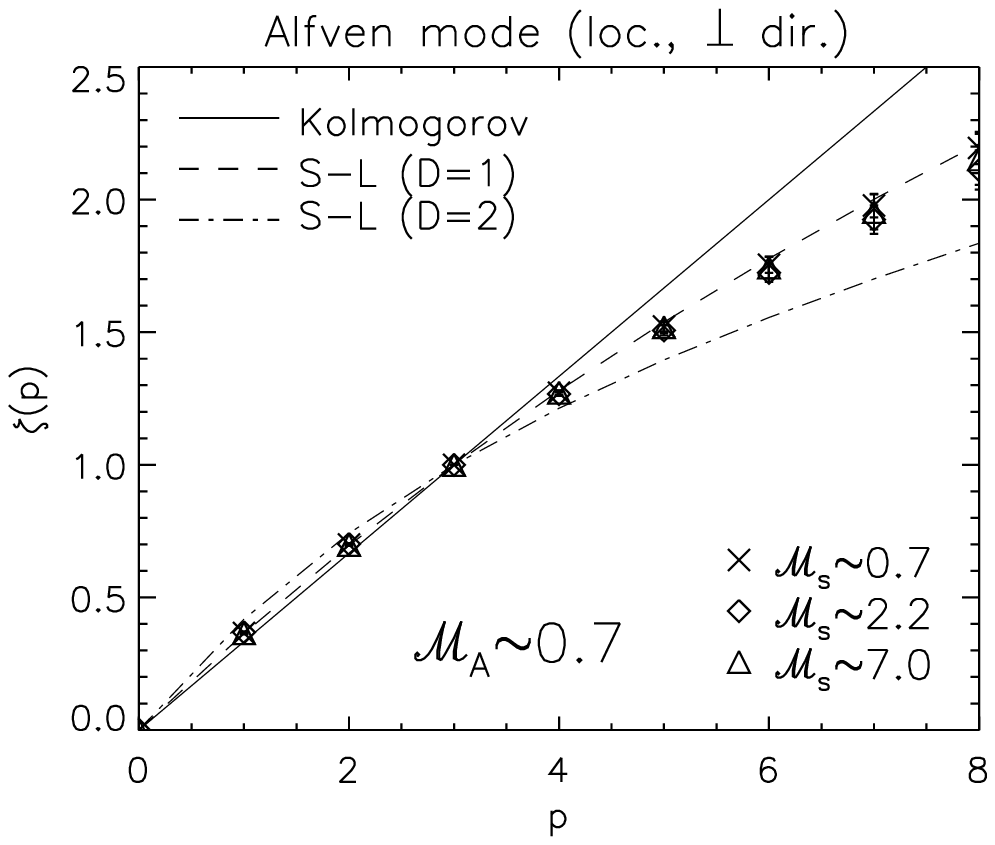}
 \plotone{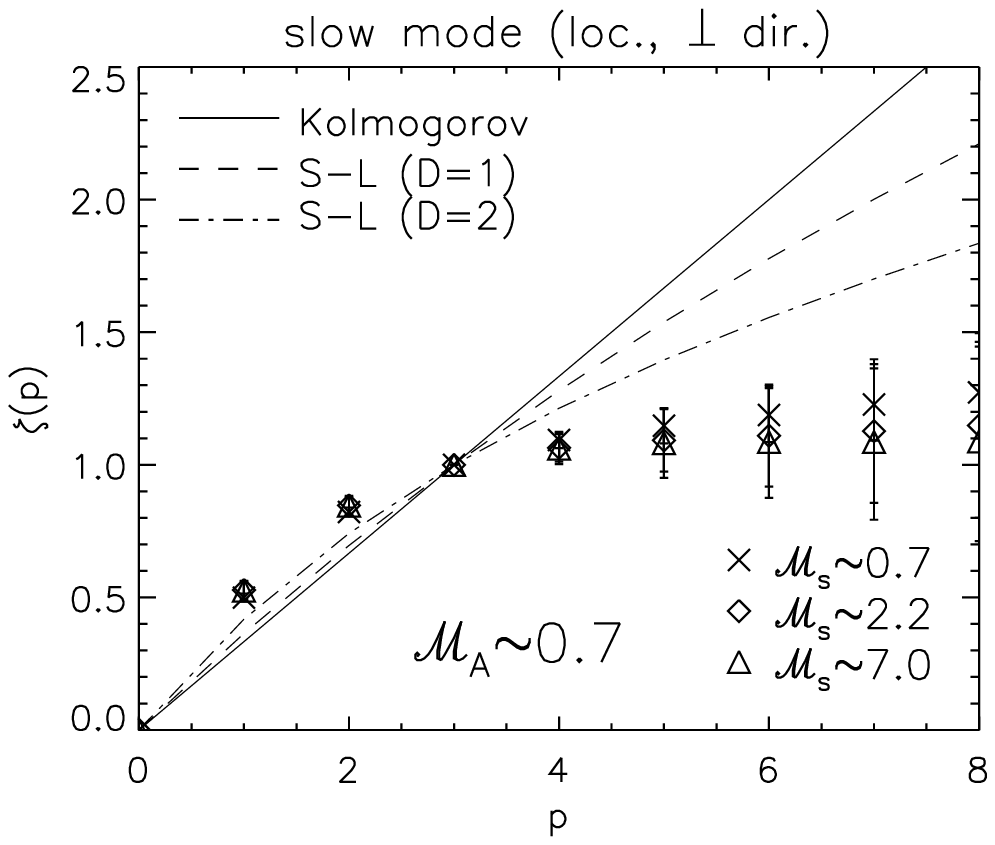}
 \plotone{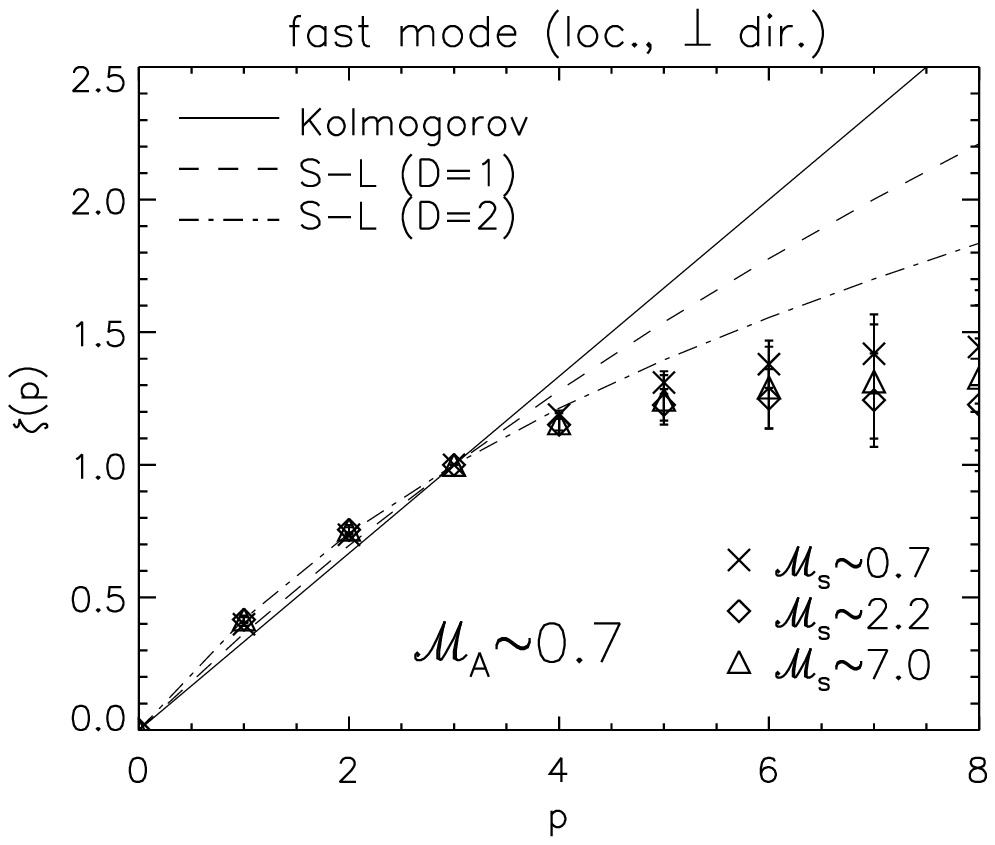}
 \caption{Scaling exponents of the Alfv\'{e}n, slow and fast waves of the velocity (left, middle and right columns, respectively) calculated in the local reference frame. Two rows show the scalings exponent at the parallel and perpendicular direction the local mean magnetic field for models with a strong magnetic field. \label{fig:expons_waves_local}}
\end{figure*}

The above considerations were carried for the scalings relations in the global
reference frame, i.e., when we do not take the direction of the local magnetic
field into account.  The strong magnetic field is dynamically important in
subAlfv\'{e}nic turbulence and can greatly influence the generation of the
structure, thus the statistical methods which include also its local direction
in the analysis can give a substantial insight into the physics of turbulence
with the presence of the magnetic field.  In Figure~\ref{fig:expons_parts_local}
we show scaling exponents calculated in the local reference frame for velocity
and its incompressible and compressible parts.  Starting from the
subAlfv\'{e}nic turbulence, which are shown on two upper rows (for parallel and
perpendicular directions, respectively), we see that the scaling exponents are
different depending on the direction.  For example, the structures of velocity
in the direction perpendicular to the local field are one-dimensional, as
suggested by the plot in the second row on the left in
Figure~\ref{fig:expons_parts_local}.  This scalings are very similar to those
calculated in the global reference frame.  However, the same velocity field in
the direction along the local magnetic field have different structures with
higher dimensions than two, according to the corresponding plots.  Both
scalings, in parallel and perpendicular directions, depend on the sonic Mach
number only marginally.  Next, the solenoidal component shown in the middle
column of figure~\ref{fig:expons_parts_local} have a similar property.  Its
structures in the perpendicular direction are very similar to that observed in
the global reference frame while in the parallel direction its structures are
two-dimensional.  Again, the potential component, although its scaling relation
cannot be explained by the current theoretical models, have scaling relations in
the perpendicular direction very similar to that in the global reference frame.
All these observations signify that the dominant structures are created in the
directions perpendicular to the local field, while the parallel structures are
less significant and usually have more dimensions.  We see that in these models
the role of the magnetic field in generation of the structure is very clear.
What do we expect when the magnetic field is weaker in such turbulence.

Now, we compare the scaling exponents of MHD waves obtained in the local
reference frame which are presented in Figure~\ref{fig:expons_waves_local}.  The
Alfv\'{e}n wave which is incompressible and directed perpendicularly to the
local magnetic field is presented in the left column of
Figure~\ref{fig:expons_waves_local}.  In the subAlfv\'{e}nic turbulence when the
magnetic field is strong we expect that the most of the structure of the
Alfv\'{e}n mode should be generated in the direction perpendicular to the local
magnetic field.  This is confirmed by the corresponding plots in
Figure~\ref{fig:expons_waves_local}.  The scaling exponents in the perpendicular
direction are consistent with those in the global reference frame
(Figure~\ref{fig:expons_waves}).  The scalings suggest the one-dimensional
dissipative structures in all subAlfv\'{e}nic models independently of the sonic
Mach number.  In the parallel direction the plot suggests the dissipative
structures with higher direction but at the same time we see very large error
bars which signify very poor statistics.  It means that the structures of the
Alfv\'{e}n wave created in the direction parallel to the local field are
marginal.  On the contrary, the slow wave has direction parallel to the local
mean magnetic field, what signifies that its dissipative structure should be
more dominant in the parallel direction.  Indeed, the slow wave in the parallel
direction has structures consistent with those observed in the global mean
magnetic field while in the perpendicular direction it shows some random, rare
events in the structure.

\section{Discussion}
\label{sec:discussion}

\subsection{Major Accomplishments and Limitations of the Present Study}

In the paper we have introduced a new procedure of decomposition of MHD
turbulence field into Alfv\'en, slow and fast modes which uses wavelets.
Compared to the decomposition procedure based on Fourier transforms described in
CL03, the wavelet decomposition is more local, thus it follows better the local
magnetic field direction in respect to which the decomposition into modes takes
place.  As a result, we expect that the wavelet decomposition procedure to be
more accurate for larger amplitudes of turbulence, i.e. larger perturbations of
magnetic field.

Our decomposition of the MHD turbulence confirmed the results in CL03 in terms
of spectra, namely, that the Alfv\'enic and slow modes are anisotropic and
consistent with the predictions of the GL95 model of incompressible turbulence,
while the fast modes are mostly isotropic and form an acoustic turbulence
cascade \citep{lithwick01,cho02a}.  As these results were used in studies of
cosmic ray scattering and acceleration
\citep{yan02,yan04a,yan08,cho05,brunetti07} as well as charged dust acceleration
\citep{lazarian02,yan03,yan04b,yan09} this is an encouraging development.

At the same time, the intermittency of the different MHD modes were shown to be
very different.  We clearly see the dependence of high order statistics of
compressible motions on the Mach number.  We interpret this dependence as the
result of shock formation, which eventually changes the nature of the
compressible motion cascade compared to the CL03 assumptions.

The limitations of the present study arise from the yet unclear nature of the
turbulent cascade.  For instance, it was shown in \cite{beresnyak09} that the
degree of locality of interactions in hydrodynamic and MHD cascade are
different.  Thus even largest available MHD simulations may not present the
actual inertial range of the cascade, but the measured slope may be strongly
affected by the extended bottle-neck effect of the simulations.  In addition,
the limited range over which Alfvenic turbulence is weak may exhibit a rather
different scaling of fast modes as a result of the interactions of the Alfvenic
and fast modes \citep{chandran05}.

In addition, within the present study we intentionally do not consider the
scaling of magnetic perturbations.  The velocity and magnetic perturbations for
subAlfvenic turbulence show some differences, which are rather difficult to
study reliably with the available numerical simulations.  These differences are
not a part of the GS95 picture, but may reflect additional yet unclear
properties of the MHD cascade \cite[see][]{mueller00}.

In our study we used only the incompressible driving.  In the presence of the
compressible supersonic driving \citep{federrath09} the scaling looks different,
but the existence of the inertial range is then questionable.  \cite{kritsuk09}
claimed that combining the compressible and incompressible driving in the Mach
number dependent fashion one can obtain a better power-law inertial range.  This
issue requires further studies.

The turbulence driving in our study is balanced, in the sense that the energy
flows in opposite directions are equal.  In the presence of sources and sinks of
turbulent energy, astrophysical turbulence is expected to be imbalanced.  Our
numerical studies of imbalanced turbulence in \cite{beresnyak10} show that the
properties of Alfvenic turbulence changes substantially in the presence of
imbalance.  However, the degree of sustainable imbalance in compressible
turbulence is still unclear.  One expects the density fluctuation in turbulent
fluid to reflect the incoming waves, altering the imbalance.  We believe that in
high Mach number fluids the imbalance is low due to the existence of substantial
density contrasts.

\subsection{Astrophysical Implications of the Turbulence Anisotropy and Intermittency}

Depending on driving astrophysical turbulence may be subAlfv\'enic, if the
injection velocity $V_L$ is less than Alfven speed $V_A$, Alfvenic, if
$V_L=V_A$, and superAlfvenic, if $V_L>V_A$.  This frequently is also described
by the Alfven Mach number $M_A=V_L/V_A$.  Formally, the GS95 model applies only
to incompressible motions with $V_L=V_A$, or equivalently $M_A=1$.  Some of the
astrophysical applications of the model, indeed, use the original form of the
theory, which substantially limits the applications of the theory
\cite[see][]{narayan01}.  However, the model can be easily generalized to cover
extensive ranges of superAlfvenic and subAlfvenic turbulence
\cite[see][]{lazarian99,lazarian06a}.  For subAlfvenic turbulence with isotropic
driving at the scale $L$ an initial weak cascade, in which the parallel scale of
motions stays the same and the spectrum $E(k_{\bot})\sim k_{\bot}^{-2}$ is
applicable, transfers to the regime of strong turbulence at the scale of
$LM_A^2$, for which the GS95 critical balance arguments are applicable.  For
superAlfvenic turbulence, while up to the scale $LM_A^{-3}$ the turbulence is
hydrodynamic, it approaches the GS95-type regime for smaller scales. Therefore,
the relations obtained for MHD turbulence that we have studied above can be
generalized for cases of different intensity of driving.

While the GS95 model is a model of incompressible turbulence, our simulations
confirm the numerical findings in \cite{cho02a} and CL03 that the scaling of the
Alfvenic mode in the compressible turbulence is very similar to its scaling in
the incompressible case.  In particular, the GS95 anisotropy of MHD turbulence
determines the rate of magnetic field wandering which is important for many
astrophysical processes, including the ubiquitous process of magnetic
reconnection \citep{lazarian99}.  Additional implications of magnetic field
wandering include the diffusion of heat and cosmic rays, MHD acceleration of
dust etc. \cite[see][for a review]{lazarian09b}.  The wavelet approach has the
potential of increasing accuracy while studying small-scale anisotropy in
simulations with strongly perturbed magnetic fields.

\cite{falgarone05,falgarone06,falgarone07} and collaborators \cite[and refs.
therein]{hily-blant07a,hily-blant07b} attracted the attention of the
interstellar community to the potential important implications of intermittency.
 A small and transient volume with high temperatures or violent turbulence can
have significant effects on the net rates of processes within the ISM.  For
instance, many interstellar chemical reactions (e.g., the strongly endothermic
formation of CH$^+$) might take place within very intensive intermittent
vortices.  The aforementioned authors claimed the existence of the observational
evidence for such reactions and heating, but a more quantitative approach to the
problem is possible. \cite{beresnyak07} \cite[see also][]{lazarian09b} used the
intermittency scaling and calculated the distribution of the dissipation rate in
the turbulent volumes.  In doing so they used the fact that \cite{she94} model
of intermittency corresponds generalized log-Poisson distribution of the local
dissipation rates \cite{dubrulle94,she95}.  The obtained rates of enhancement
were not sufficient to explain the heating required for inducing interstellar
chemistry \citep{beresnyak07}.  The same approach was used by \cite{pan09} who
obtained, however, a different result. We believe that one should distinguish
shocks from vortical motions while calculating the heating induced by
intermittency.  Our present study show very different scalings relevant to these
types of motions.

\subsection{Studies of Compressible MHD Turbulence in Astrophysical Context}

Numerous studies of compressible MHD turbulence are done in the context of star
formation \cite[see reviews by][and references therein]{maclow04b,mckee07}.
Most of these simulations are focused on the large-scale appearances of the
turbulence, which is determined by the turbulent driving and do not exhibit any
extended inertial range of turbulence.

Search for the universal relations for compressible turbulence resulted in the
rise of interest to the \cite{fleck83} idea of searching universality not for
velocity, but for the combination of the velocity and density in the form
$\rho^{1/3} v$.  The numerical study of hydrodynamic compressible turbulence
revealed that, indeed, the density modified velocity shows the same Kolmogorov
scaling both for low and high Mach number turbulence \citep{kritsuk07}.  Similar
effect was confirmed in our MHD simulations \cite{kowal07b}.  However, the
physical justification of this universality is unclear and it may result just
from the coincidental compensation of the change of velocity and density indexes
as shocks develop at high Mach number turbulence.

We report steepening of the spectra of compressible motions at high Mach
numbers.  High resolution hydro simulations \cite[see][]{kritsuk07} show that
the velocity spectrum becomes steeper for high Mach number simulations.  This
corresponds to the observational studies of the supersonic velocity turbulence
in \cite{padoan06,padoan09} and \cite{chepurnov06}.  These studies are done with
the Velocity Channel Analysis (VCA) and Velocity Coordinate Spectrum (VCS)
techniques, which are theory-motivated and tested techniques
\citep{lazarian00,lazarian04,lazarian06b,lazarian08,chepurnov08}.  The
application of these techniques should enhance the range of astrophysical
turbulent velocity fields that can be studied observationally\footnote{Recent
examples of the techniques of observational studies of the turbulent density
field can be found in \cite{kowal07a,burkhart09,burkhart10}. The velocity is a
more covered statistics, but it is more difficult to study \cite[see][for a
review]{lazarian09a}.}  It is comparing numerics, observations and theory that
the progress in understanding of turbulence requires.

\section{Summary}
\label{sec:summary}

In this article we presented a new technique of decomposition of turbulent MHD
motions into Alfven, slow and fast modes. The technique is based on the use of
wavelets, which provide a more local decomposition compared to the Fourier
approach in CL02 and CL03. This enables one to have better accuracy of the
decomposition of MHD turbulence into fundamental modes for higher amplitude of
magnetic perturbations.

By applying the wavelet decomposition to the results of our simulations of
compressible MHD turbulence, we investigated the scaling properties of velocity
in compressible MHD turbulence for different sonic $M_s$ and Alfv\'{e}nic $M_A$
Mach numbers. We analyzed spectra, the anisotropy, scaling exponents and
intermittency of the total velocity and its components corresponding to the
Alfv\'{e}n, slow and fast modes. We found that:
\begin{itemize}
 \item The amplitude of velocity fluctuations depends on $M_s$ only marginally.
The lack of significant dependence of the velocity fluctuations is also observed
for its incompressible part, as well as for the Alfv\'{e}n and slow waves. The
compressible part of velocity and the fast wave show a dependence on ${\cal
M}_s$, but only for subsonic turbulence. In the case of supersonic models, the
fluctuations of the compressible part and fast mode of the velocity have
comparable amplitudes.

 \item The spectral indices depend on $M_A$ in turbulence with a strong magnetic
field. In the case of turbulence with a weak magnetic field only the indices of
spectra of the fast wave change between sub- and supersonic models. For the
other components, the spectral indices do not change appreciably with the sonic
Mach number. While our conclusions about spectra of fast modes for subsonic
turbulence agree with the CL03 conclusion about the acoustic cascade of these
modes, we feel that for high Mach number we get the spectrum of shocks. The
anisotropy of Alfv\'enic turbulence and slow modes is in agreement with GS95
theory for both the cases of high and low beta plasmas. The velocity
fluctuations of the fast modes demonstrate isotropy.

 \item In the global reference frame, we observe stronger changes of the scaling
exponents and intermittency for velocity and its all components with ${\cal
M}_s$ in the case of turbulence with a weak magnetic field. The intermittency of
structures grows with the values of ${\cal M}_s$. However, when the external
magnetic field is strong, the intermittency for all components depends on the
sonic Mach number only marginally. In the local reference frame, the scaling
exponents turbulence depend on the direction with respect to the direction of
the local mean magnetic field. The dependence is stronger for the
subAlfv\'{e}nic turbulence.
\end{itemize}

\acknowledgments
Our research is supported by the NSF grant AST-0808118 and the Center for
Magnetic Self-Organization in Astrophysical and Laboratory Plasmas.


\end{document}